\documentclass[12pt]{article}
\usepackage{graphicx}
\usepackage{amssymb}
\usepackage{epstopdf}
\newcommand{\ignore}[1]{}
\newcommand{\be}{\begin{equation}} \newcommand{\ee}{\end{equation}}
\newcommand{\ba}{\begin{eqnarray}} \newcommand{\ea}{\end{eqnarray}}
\newcommand{\nn}{\nonumber} \renewcommand{\bf}{\textbf}
\newcommand{\ra}{\rightarrow}

\renewcommand{\a}{\alpha} \renewcommand{\b}{\beta}

\newcommand{\p}{\partial}

\def\slashb#1{\setbox0=\hbox{$#1$}#1\hskip-\wd0\dimen0=5pt\advance
        \dimen0 by-\ht0\advance\dimen0 by\dp0\lower0.5\dimen0\hbox
          to\wd0{\hss\sl/\/\hss}}
\def\bra#1{\left< #1\right|}
\def\ket#1{\left| #1\right>}
\def\bracket#1#2{\left<#1\mid #2\right>}
\def\EV#1#2#3{\bra{#1}#2\ket{#3}}
\input{epsf}

\begin{document}
\bibliographystyle{apj}

\title{Testing Isotropy of \\ Cosmic Microwave 
Background Radiation }
\author{ Pramoda Kumar Samal$^{1}$, Rajib Saha$^{1}$ Pankaj~Jain$^{1}$ 
\\   John~P.~Ralston$^2$} 
\maketitle

\begin{center}
{$^{1}$Department of Physics, Indian Institute of Technology, Kanpur, U.P, 208016, India \\ }
{$^{2}$Department of Physics \& Astronomy,
University of Kansas, Lawrence, KS 66045, USA\\}
\end{center}

\begin{abstract}
We introduce new symmetry-based methods to test for isotropy in cosmic microwave 
background  radiation.  Each angular multipole is factored into unique products of 
power eigenvectors, related multipoles and singular values that provide 2 new rotationally invariant measures mode by mode. The {\it power entropy} and {\it directional entropy} are new tests of randomness that are independent of the usual CMB power.  Simulated galactic plane contamination is readily identified, and the new procedures mesh perfectly with linear transformations employed for windowed-sky analysis.  The ILC -WMAP data maps show 7 axes well aligned with one another and the direction Virgo.  Parameter free statistics find 12 independent cases of extraordinary axial alignment, low power entropy, or both having 5\%  probability or lower in an isotropic distribution.  Isotropy of the ILC maps is ruled out to confidence levels of better than 99.9\%, whether or not coincidences with other puzzles coming from the Virgo axis are included. 
Our work shows that anisotropy is 
 not confined to the low l region, but extends over a much larger $l$ range. 
\end{abstract}


\section{Introduction}

Studies of the cosmic microwave background (CMB) were long posed in terms of the temperature power spectrum. The power spectrum is invariant under rotations, and by itself cannot in principle test the basic postulate that the radiation field should be statistically isotropic. 
The availability of high quality data from WMAP (Hinshaw {\it et al} 
2003, 2007) has allowed isotropy to be examined critically, and with surprising outcomes.  

A long-standing question exists in the unexpected size of low-$l$ multipoles. Interpretation of this is stalemated by inherent uncertainties of fluctuations called cosmic variance. Directional effects are much more decisive, because they confront a symmetry.  de Oliveira-Costa, et al (2004) constructed an axial statistic for which they found modest statistical significance in multipoles for $l= 2,\,  3 $ being rather well aligned.  The fact that the dipole ($l=1$) also aligns very closely with the multipoles $l= 2,\,  3 $ was later highlighted by Ralston and Jain (2004) and Schwarz {\it et al} (2004). When the dipole is interpreted to be due to our proper motion, it has sometimes been excluded as having ``no cosmological significance.''  However there are many physical mechanisms which can correlate CMB observations with galactic motion.  There are good reasons not to exclude it, and the correlation of all three multipoles significantly aligned with the constellation Virgo is
  quite inconsistent with chance. 
  
Subsequently there have been a large number 
  of studies (Eriksen {\it et al} 2004, Katz and Weeks 2004, Bielewicz 
  {\it et al} 2004, Bielewicz {\it et al} 2005,
  Prunet {\it et al} 2005, Copi {\it et al} 2006, 
  de Oliveira-Costa and Tegmark 2006, Wiaux {\it et al} 2006, 
  Bernui {\it et al} 2006, Freeman {\it et al} 2006, Magueijo and  Sorkin 2007, Bernui {\it et al} 2007,
  Copi {\it et al} 2007,  Helling {\it et al} 2007, Land and Magueijo 2007) which claim the CMB is not consistent with isotropy. Several physical explanations for the 
  observed anisotropy have been put forward (Armendariz-Picon 2004, Moffat 2005, Gordon {\it et al} 2005, Vale 2005, Abramo {\it et al} 2006, Land and Magueijo 2006, Rakic {\it et al} 2006
Gumrukcuoglu {\it et al} 2006, Inoue and Silk 2006, Rodrigues 2007, 
Naselsky {\it et al} 2007, Campanelli {\it et al} 2007). 
It has also been suggested that the anisotropy may be due to foreground 
contamination (Slosar and Seljak 2004).
Land and Magueijo (2005) find evidence that the detected anisotropy has
positive mirror parity.
Meanwhile some studies find no inconsistency ( Hajian {\it et al} 2004, 
Hajian and Souradeep 2006, Donoghue and  Donoghue 2005). 
There have also been several studies of the primordial perturbations 
(Koivisto and Mota 2006,  Battye  and  Moss 2006, Armendariz-Picon 2006, Pereira {\it et al} 2007, Gumrukcuoglu {\it et al} 2007) and inflation (Hunt and Sarkar 2004, Buniy {\it et al} 2006) in an anisotropic universe. 
  
  It is clearly necessary to explore new observables to determine whether anisotropy may be signs of physics beyond the standard paradigm. 
Here we develop new methods to test CMB data for isotropy.  The new tests are possible because there exist many more invariant and vector-valued quantities 
than commonly examined. As conventional, the temperature distribution $\Delta T(\hat n)$ is expanded in terms of the spherical harmonics, defining $\Delta T(\hat n)=\sum_{lm}a_{lm}Y_{lm}(\hat n)$, where $\hat n$ is a unit vector on the sky. The assumption of statistical isotropy is the statement that the ensemble average is given by \ba <\, a_{lm}a_{l'm'}^{*} \,> =C_{l}\delta_{ll'}\delta_{mm'}. \label{default} \ea As a consequence all tensors formed from products of $a_{lm}$ must be isotropic.  We concentrate on the second rank tensors that can be formed, which include the {\it isotropic power} $C_{l} \, \delta_{ij} $ and the {\it power tensor} $A_{ij}(l) $.  They are defined by \ba A(l) \, \delta_{ij} ={1 \over 2 l+1}\sum_{mm'} \, a_{lm}^{*}\delta_{ij}\delta_{mm'}a_{lm'}; \nn  \\ 
A_{ij}(l) ={1 \over l(l+1)} \sum_{mm'} \, a_{lm}^{*}(\, J_{i}J_{j} \,)_{mm'}a_{lm'}. \ea Here $J_{i}$ are the angular momentum operators in representation $l$ for Cartesian indices $i=1...3$.  From Eq. \ref{default} statistical isotropy predicts the ensemble averages \ba <\, A   \,>  = C_{l} ; \nn \\ <\, A_{ij}(l)   \,>  = C_{l}\delta_{ij}. \ea  

Giving attention to the power tensor represents a modest step towards examining the data more thoroughly.  Consider the region of $0<l<300$, which includes the acclaimed peak in plots of $l(l+1)C_{l}$. There are 299 data points from $2 \leq l \leq 300$ but the information makes a smooth curve, one that can be fit 
with 3 or 4 parameters.  The existence of an acoustic peak is useful but only the beginning of tests.  Each $a_{lm}$ has $2l+1$ components, and the remaining 90,298 numbers, measured at great effort and considerable public expense, are never examined with the power spectrum. One reason comes from early tradition when data was poor.  There also appears to be a widespread but false belief that the power spectrum is the only rotationally invariant quantity.  

The approach of Copi {\it et al} (2007) consists of factoring the spherical harmonics expansion into products of vector-valued terms. All spherical harmonics can be developed from traceless symmetric polynomials of vector products. 
There are $2l+1$ real freedoms in multipoles of given $l$.  Each mutipole can be factored into $l$ unit vectors, each with 2 freedoms, making $2l$ terms
(Weeks 2004, Katz and Weeks 2004).  Finally there is one overall power coefficient. The analysis of 50 vectors for $l=50$ (say) is inherently complicated, and complicated further by discovery that Maxwell-multipole vectors of isotropic data are kinematically correlated to ``repel'' one another. Dennis (2005) discusses this and even solves the 2-point correlation analytically.   
 
Factoring into products of vectors is natural for tensors of rank 2. When we consider rank-2 tensors, we realize they have more than one invariant eigenvalue, so that the CMB power cannot even describe the quadrupole component.  Tensors of rank higher than 2 have many types of invariants.  One complication in finding 
them is the absence of a unique canonical form for high-rank tensors (da Silva and Lim 2006). Another technical obstacle is to obtain the invariants using quadratic functions of the data, which are statistically robust. 

The technical problems are solved here with a procedure exploiting an invariant canonical form where {\it unique and preferred vectors} are factored from each multipole. The fact this can be done is both obvious from symmetry and systematic via Clebsch-Gordan series.  It is a particular {\it linear transformation} of the $a_{lm}$ that leads to the power tensor $A_{ij}$ as quadratic form. When our procedure is applied to WMAP data there are numerous new invariants and several new signals of anisotropy.  

We had two distinct motivations in beginning our study: \begin{itemize} 
\item Isotropy is a well-defined statistical data question that does not need any physical hypothesis to be motivated.  Physical hypotheses are best characterized and tested through their symmetries.  No free parameters are used in expressing isotropy, which can be confirmed or ruled out independent of the parameterization of model predictions. All of the new tests in this paper are also parameter-free.  Isotropy of the CMB is also a separate issue from isotropy of other observables in the universe.  This is because the CMB contains both information on formation in interaction with matter, and information on the propagation of light since the era of decoupling. By separating tests of isotropy of the CMB we maintain an orderly process of testing larger issues in cosmology.

\item Anisotropy of the CMB may be a signal of physics beyond electrodynamics and gravity. The interaction of light with an anisotropic axion-like condensate, 
or its conversion into weakly interacting particles, may give anisotropic signatures without disturbing the large scale distribution of matter.  

Interestingly, there is a long history of ``Virgo alignment'' of electromagnetic propagation effects which cannot be explained by any known media of conventional physics. Propagation of radiation in the sub-GHz region of radio telescopes, the 20-90 GHz region of CMB studies, and the optical region all reveal anisotropic effects. Many years ago Birch (1982) observed that the offsets of sub-GHz linear polarizations relative to radio galaxy axes were not distributed isotropically. The statistical significance of Birch's signal was confirmed by Kendall and Young (1984) and a statistic used by Bietenholz and Kronberg (1984), but generally rejected nonetheless.  The redshift-dependent signal found by Nodland and Ralston (1997) was also generally dismissed.  Rejection was based on not finding correlation in statistics believed to test for the same signals.  Yet one must be sure the same tests were really made. Paradoxes were finally resolved by classification under {\it parity}. Statistics finding no signal test for 
{\it even parity} distribution with no sense of twist, while those statistics of {\it odd parity} consistently show signals of high statistical significance.  Subsequent analysis (Jain and Ralston 1999, Jain and Sarala, 2006) has found a robust axial relationship of {\it odd parity} character and aligned with Virgo.   

Until recently these explorations of anisotropy were conducted in the radio regime, which might indicate a conventional explanation in plasma electrodynamics.  The discovery of {\it optical frequency} polarization correlations (Hutsem\'{e}kers (1998), Hutsem\'{e}kers and Lamy (2001)) from cosmologically distant QSO's is extraordinary. The optical correlations again generate an axis aligned with Virgo (Ralston and Jain 2004). The coincidence of 5 closely aligned axes coming from 3 distinct frequency domains makes an exceptional mystery that merits further study.   

\end{itemize}

In this paper we make no attempt at physical explanation of anisotropy. Our new methods provide two independent new roads for data analysis.  Order by order for each $l \geq 2$ there exist three rotationally invariant eigenvalues of 
$A_{ij}(l)$.   The sum of three eigenvalues reproduce the usual power $C_{l}$. Meanwhile 
two independent combinations are new invariants.  For the region of $2\leq l \leq 50$, say, we have 98 new invariants never examined before.  The isotropic CMB model predicts that all three eigenvalues should be degenerate and equal to $C_{l}/3$, up to fluctuations.  We test this prediction by introducing the {\it power entropy} of $A_{ij}$.   
Inspired by quantum statistical mechanics, the power entropy is a stable and bounded measure of the randomness of eigenvalues of matrix $A_{ij}$.  The entropy of the data turns out not to be consistent with the conventional CMB prediction.  Next we examine the eigenvectors of $A_{ij}$.  The eigenvectors are independent {\it covariant} quantities. Under the isotropic prediction they should be randomly oriented.  For the range of $2\leq l \leq 50$ there are 98 independent eigenvectors, and 49 ``principal'' eigenvectors associated with the largest eigenvalues.  The largest eigenvalue makes the largest contribution to the total power and gives an objective reason to be singled out. Statistical measures of the alignment of eigenvectors do not happen to support the random isotropic expectations. Instead 6 axes obtained from the range $2\leq l \leq 50$ are found to be well aligned with the dipole or Virgo axis, each case having independent probability of less than $5\%$. 

\subsection{Outline of the Paper}

Transformation properties and signal processing of $a_{lm}$ are encoded by linear operations.  To take advantage we first introduce $\psi_{m}^{k}$, which is the unique linear transformation of $a_{lm}$ that divides each $l$-multiplet into direct products of irreducible representations with vectors.  We relate $\psi_{m}^{k}$ to the power tensor, and study the distribution of invariants in the ILC map. 

Complementary to these studies are comparisons with ``masked-sky'' power spectra, evaluated with a window-function blocking out a particular angular region.  Our linear transformations are made to mesh perfectly with the linear operations of window functions and subtractions.  In studies we report, we sometimes masked out the plane of the galaxy, and divided the remainder into two ``top'' and ``bottom'' regions.  We study statistics of the regions as given, and also the statistics of full-sky maps they predict under standard procedures. There are several reasons for the masked sky studies. First, we can both study top/bottom asymmetries as a simple probe of isotropy {\it including} the galactic plane.  We also study ``cone-masks'' which retain only a region subtended by cones at high galactic latitudes, reducing the importance of galactic plane contamination.  Finally we make simulations of galactic contamination and show that our new methods readily detect contamination and 
 find its angular location.  The ILC data maps turn out to show significant anisotropy that cannot be attributed to galactic foregrounds.

In the following section we describe the new methods.  
Applications to data follow. Not surprisingly, the power tensor method is generally more sensitive to anisotropy than power spectrum methods. Yet power 
spectrum methods find similar signals of anisotropy. This is significant because the two approaches are statistically independent - nothing about the power 
spectrum is used in the new approach.  Almost all of our studies develop new information that by construction is independent of the usual $CMB$ power $C_{l}$. 

The reliability of CMB maps is not under our control and certainly open to debate. The observers of the ILC team present maps claimed to have systematic errors equal to or smaller than cosmic variance.  For this reason we don't discuss systematic errors.  Our results are quantified using statistical errors based on Monte Carlo simulations with random realizations in place of the $a_{lm}$. 

\section{Covariant Frames and Invariant Singular Values}

In this Section we use Dirac notation with \ba a_{lm}=\bracket{l,\,m}{a(l)}. \nn \ea (There should be no confusion with ensemble averages.)  The states $\ket{lm}$ are eigenvectors of angular momentum $\vec J_{z}$ and $\vec J^{2}$ with eigenvalues $m, \, l(l+1)$.  The temperature measurements are real valued, which produces a constraint in the usual basis convention: 
\begin{eqnarray}
a_{lm}=(-1)^m \, a_{l-m}^* \nn
\end{eqnarray}  Although the phase conventions standardize the $a_{lm}$ they have no effect on anything we report. 

Besides simplicity, the power tensor is well motivated on rotational symmetry grounds. Each multipole $a_{lm} \ra \ket{a(l)}$ represents a geometrical object visualized as a complicated beach-ball.  ``Holding by hand'' the object for inspection, make a small rotation by angle $\vec \theta$, under which \ba  \ket{a(l)} \ra  \ket{a(l)}'= \ket{a(l)} + \ket{\delta a(l)} ; \nn \\ 
 \ket{\delta a(l)}= -i\vec \theta\cdot \vec J  \ket{a(l)}. \nn \ea  Under what axes does the rotation make the most difference ?  Compute the Hessian of the change, \ba {\p^{2}  \over \p \theta_{i} \p\theta_{j}}\bracket {\delta a(l)}{\delta a(l)}  &=&  \EV{a(l)}{J^{i}J^{j}}{a(l)} \nn \\  &=& A^{ij} \nn  \ea By Rayleigh-Ritz variation, the maximum rotation is developed along the eigenvectors of  $A^{ij} $, which are natural {\it principal axes} of the $a_{lm}$ as objects order by order in $l$.

It is convenient to define a linear map or ``wave function" $\psi^k_{m}(l)$ (Ralston and Jain 2004) we call ``vector factorization:''
\begin{equation}
\psi^k_{m}(l) = {1\over \sqrt{l(l+1)}} \langle l, \, m|J^k|a(l)\rangle.
\end{equation}  The purpose of this transformation is to extract (algebraically divide out) a vector (spin -1) quantity from each representation of spin$-l$, so each can be covariantly compared across different $l$. Inserting a complete 
set of states gives
\begin{eqnarray}
\psi^k_{m}(l) &=& {1\over \sqrt{l(l+1)}} \sum_{m'=-l}^l \langle l, \, m|J^k|l, \, m'\rangle \langle l, \, m'|a(l) \rangle\cr
&=&\sum_{m'=-l}^{l}\Gamma_{mm'}^k \, a_{lm'}
\end{eqnarray}
where 
\begin{equation}
 \Gamma_{mm'}^k = {1\over \sqrt{l(l+1)}} \langle l, \, m|J^k|l, \, m'\rangle \nn 
\end{equation}  Under rotations each index of $\psi_{m}^{k}$ rotates by its representation $R(j)$, namely \ba \psi_{m}^{k} \ra \psi_{m}^{k'} = R^{kk'}(1)R_{mm'}(l)\psi_{m}^{k} . \label{rot} \ea 
The transformation $a \ra \psi$ is invertible.  The inverse relation can be written \ba  a_{lm} =  \sum_{km' } \, \Gamma_{k}^{mm'}\psi_{m_{}}^{k}, \nn \ea where 
\ba \Gamma_{k}^{mm'}  &=& {1\over \sqrt{l(l+1)}} \langle l, \, m|J^k|l, \, m'\rangle, \nn \\ &=& \Gamma_{mm'}^k .\nn \ea Upon rotating $\psi$ by the rule of Eq. \ref{rot}, then \ba a_{lm} \ra a_{lm}'= \Gamma_{k}^{mm'} R^{kk'}(1)R_{mm'}(l)\psi_{m}^{k}, \nn \\ = R_{mm'}(l)a_{lm'}.  \nn \ea The standard power spectrum invariant is {\it one} of the invariants produced from $\psi_{m}^{k}$: it is \ba \sum_{mk} \,  \psi_{m}^{k} \psi_{m}^{k*}=\sum_{m} \, a_{lm}a_{lm}^{*}. \nn \ea 

Mathematically $ \psi$ represents the multiplet $a$ as a {\it unique} sum of outer products of vectors $e^{\a}$  (spin-1 representation) with representations $u^{\a}$ of spin$-l$.  The terms are obtained by the singular value decomposition (SVD),
\begin{equation}
\psi^k_{m}(l) = \sum_{\alpha=1}^3 \,  e_k^{\alpha} \Lambda^{\alpha} u_{m}^{\alpha *}
\end{equation}  Here and in subsequent equations we suppress label $l$ when it is obvious. The Appendix discusses factorization in general terms of a Clebsh-Gordan series. 

The singular values $\Lambda^\alpha$ are invariants under rotations on the indices $k$ and $m$. (Indeed they are invariants under the even higher symmetry of {\it independent} $SO(3) \times SO(3)$ rotations.) The various factors are constructed by diagonalizing the $3\times 3$ Hermitian matrices 
\ba 
(\psi \psi^{\dagger})^{kk'}(l) =  \sum_m \,  \psi_{m}^k(l) \psi_{m}^{k'*}(l) , \nn \\
 =\sum_{\a}\, e_{i}^{\a } ( \Lambda^{\a})^{2}  e_{i}^{\a *} ; \label{emade} \\
  (\psi^{\dagger}\psi)_{mm'}(l) =  \sum_k \, \psi_{m}^{k*}(l) \psi_{m'}^{k}(l) , \nn \\
 =\sum_{\a}\, u_{m}^{\a } ( \Lambda^{\a})^{2}  u_{m'}^{\a * } . \nn \ea 
 
Since they are the eigenvectors of Hermitian matrices the $e^{\a}$ and $u^{\a}$ are generally orthogonal, except for exceptional degeneracies, 
and normalized to \ba  
\sum_k \, e_k^{\alpha *}  e_k^{\beta}= \delta_{\alpha\beta}; \nn \\
\sum_m \, u_m^{\alpha *} u_m^{\beta} = \delta_{\alpha\beta} .\nn 
\ea  Since $( \Lambda^{\a})^{2}$ are the eigenvalues of Hermitian matrices they are real and positive, with $\Lambda^{\a} >0$ defining the sign convention for $e^{\a}$.  Since $a_{lm}$ are real, we also have $e^{\a}$ real valued.   

The set of three orthogonal $e^{\a}$ defines the ``frame vectors'', which make a preferred frame for the vector components of $\psi$.  In that preferred frame and the frame of three $u^{\a}$, the matrix $\psi_{m}^{k}$ is diagonal with three invariant singular values $\Lambda^{\a}$.  We will call $\Lambda_{\a}\delta_{\a \b}$ the ``$SV$ matrix'', $SV$ standing for singular values.

\subsection{Isotropy}

The relation of $\psi_{m}^{k}$ to the power tensor $A_{ij}$ is simple algebra.  We have \ba A^{ij}(l) = {1 \over l(l+1)}  tr( \, J^{i}\ket{a(l)}\bra{a(l)} J^{j} ), \nn  \ea where $tr$ is the trace.  

Then \ba A^{ij}(l) = \sum_{m} \, \psi_{m}^{i}(l) \psi_{m}^{j*}(l) , \nn \\  =\sum_{\a}\, e_{i}^{\a }(l) ( \Lambda^{\a}(l) )^{2} e_{j}^{\a*}(l)  , \nn \ea which is just Eq. \ref{emade}.  Isotropy holds that the eigenvectors $e^{\a}$ must be distributed isotropically on the sky and that the eigenvalues $\Lambda^{\a}$ are random variables concentrated at $\sqrt{C_{l}/3}$.  In terms of the $\psi_{m}^{k}$ factors, Eq. \ref{default} predicts \ba <\, \psi_{m}^{k}(l) \psi_{m'}^{k'*}(l')  \,>={C_{l} \over 3} \delta^{ll'}\delta_{mm'}\delta^{kk'}. \label{correl} \ea 

Eq. \ref{correl} can be contrasted with the correlations of Maxwell-multipole vectors developed by dividing multipoles into products of vectors.  Those vectors have not been organized into irreducible representations nor classified in 
invariant terms of importance, and they do {\it not} have an uncorrelated distribution (Dennis 2005).  An advantage of using our $\psi_{m}^{k}$ representations is that isotropy transforms to isotropy without annoying Jacobian factors. 

Here is how to interpret the factors.  The highest possible degree of anisotropy produces one singular value equal to the total power, and two others that vanish.  The corresponding $a_{lm}$ components can be written as the product of one vector $e^{(1)}$ and one multiplet $u^{(1)}$. We will call this special situation a ``pure state.''  This has a very simple realization for the quadrupole case $l=2$.  Any quadrupole is equivalent to a symmetric $3\times 3$ matrix made from sums of products of 3-vectors. A generic symmetric matrix has 3 real and unequal eigenvalues.  If two eigenvalues vanish then the matrix is the outer product of a unique vector, a pure state.  Conversely, if all eigenvalues are degenerate, the matrix is a (trace subtracted) multiple of the unit matrix, equivalent to the sum of outer products of three frame vectors by completeness, $ 1=\sum_{\a} \, \ket{e^{\a} } \bra{e^{\a}}, $ which is the isotropic prediction.  
 
\subsection{Tensor Power Entropy}
 
The isotropy hypothesis that the $SV$ matrix should be $\sqrt{C_{l}/3}$ times the unit matrix can be tested in several different invariant ways.  
  
The {\it information entropy} or simply {\it power entropy} of the power tensor comes by recognizing that $\rho^{kk'}=(\psi \psi^{\dagger})^{kk'} $ is proportional to a density matrix on 3-space. The proportionality constant is the overall power. Removing it produces a normalized form 
\ba \tilde \rho^{kk'}={ (\psi \psi^{\dagger})^{kk'}  \over \sum_{j}\,  
(\psi \psi^{\dagger})^{jj} }. \ea  
Von Neumann (1932) found the entropy $S$ of normalized Hermitian matrix $\rho$ to be 
\ba S &=&  -tr(\, \tilde \rho \, log(\, \tilde \rho \, ) \,) ,\label{entropy} \\ &=&  \sum_\alpha \,  (\tilde \Lambda^\alpha)^2log((\tilde \Lambda^\alpha)^2),\nn \\ 0 & \leq& S \leq log(3). \nn \ea  
Here $(\tilde \Lambda^{\a})^{2}$ are normalized to sum to one, again removing the overall power from discussion: \begin{equation}
(\tilde \Lambda^\alpha)^2 = \frac{(\Lambda^\alpha)^2}{\sum_{\alpha'} \,  (\Lambda^{\alpha'})^2} \nn 
\end{equation} This normalizes $tr(\tilde \rho)=1$.  

The entropy is unique in being invariant, extensive, positive, additive for independent subsystems, and zero for pure states.  A density matrix with no information is a multiple of the unit matrix, and has entropy equalling the log of its dimension.  Very simply $S_{iso} \ra log(3)$ is the isotropic CMB prediction.  Low entropy compared to isotropy is a measure of concentration of power along one or another eigenvector of $\rho$.  In the present context pure states have all the power in one singular value, define one single directional eigenvector, and have power entropy $S_{pure} \ra 0$.  

\subsection{Power Alignments Across $l$-Classes} 

Our methods allow us to explore the alignment of power tensors between different $l$-classes. 

One of the most interesting tests concerns the ``principal axis $\tilde e(l)$'', which means the eigenvector with the largest eigenvalue for each $l$.  From Eq. \ref{correl} the set of all principal frame vectors should be a symmetrical ball in the isotropic prediction. If the data is anisotropic a bundle of vectors may lie along an axis or in a preferred plane. It is important to remember that the sign of eigenvectors is meaningless, and determined by algorithms assigning signs in computer code. Thus the ``average'' eigenvector is not a good statistic. Tensor products are the natural probe of a collection. For statistical studies we construct a matrix $X$ defined by \begin{equation}
X_{ij}(l_{max} )=\sum_{l=2}^{l_{max}}  \, \tilde e_i^l \tilde e_j^l\ .  
\end{equation}
The eigenvalues of $X$ are a probe of the shape of the bundle collected from $2 \leq  l  \leq l_{max} .$  We probe the isotropy of the collection with the {\it directional entropy} $S_{X}$ computed using Eq. \ref{entropy} and $X \ra\tilde  \rho_{X}$.  The directional entropy does not use the singular values of the CMB data, and is independent of the power entropy. Confirmation of isotropy comes if $S_{X}\sim log(3)$ up to random fluctuations.  A signal of anisotropy would be an unusually low value of $S_{X}$ compared to $log(3)$.  

\subsubsection{Traceless Power Tensor} 

We also compare alignments using a statistic that includes the weighting by singular values. 
For this purpose we define the traceless power tensor $B^{ij}$, \ba B^{ij}(l)=A^{ij}(l)-{1\over 3}tr(\, A(l)\, )\delta^{ij} .\nn \ea  The eigenvectors of $B$ are the same as $A$, but the value of the total power has been removed. To compare two angular momenta classes we examine the correlation \ba Y(l, \,l') = { tr(\,  B(l)B(l') \,) \over \sqrt{tr(\,  B(l)B(l) \,)} \sqrt{tr(\,  B(l')B(l') \,)}   }. \ea In the isotropic uncorrelated model $Y(l, \,l') $ should be proportional to $\delta_{ll'}$.

\section{Full Sky Studies}

The WMAP-ILC team developed a foreground cleaned temperature anisotropy
map using a method described as \emph{Internal Linear Combination} (ILC).  The process combines data from 5 bands of frequencies 23, 33, 41, 61, and 91 GHz. While there also exist several other cleaning procedures that are interesting to 
compare (Tegmark {\it et al} 2003, Saha {\it et al} 2006, Eriksen {\it et al}
2007), our
main focus lies on the ILC map.
 
According to Hinshaw {\it et al} (2007) the ILC map is known to become dominated by statistical instrumental noise for $l \gtrsim 400$. Errors on the $a_{lm}$ come from several sources.  Systematic errors may occur from inappropriate foreground subtractions.  Ambiguities in combining frequency bands also contribute. The range $2\leq l \leq 50$ is considered to be reliable under statements that statistical and systematic errors lie within ranges typical of cosmic variance.  We restrict our studies to this range.

\begin{figure}
\begin{center}
\includegraphics[width=4in,angle=-90]{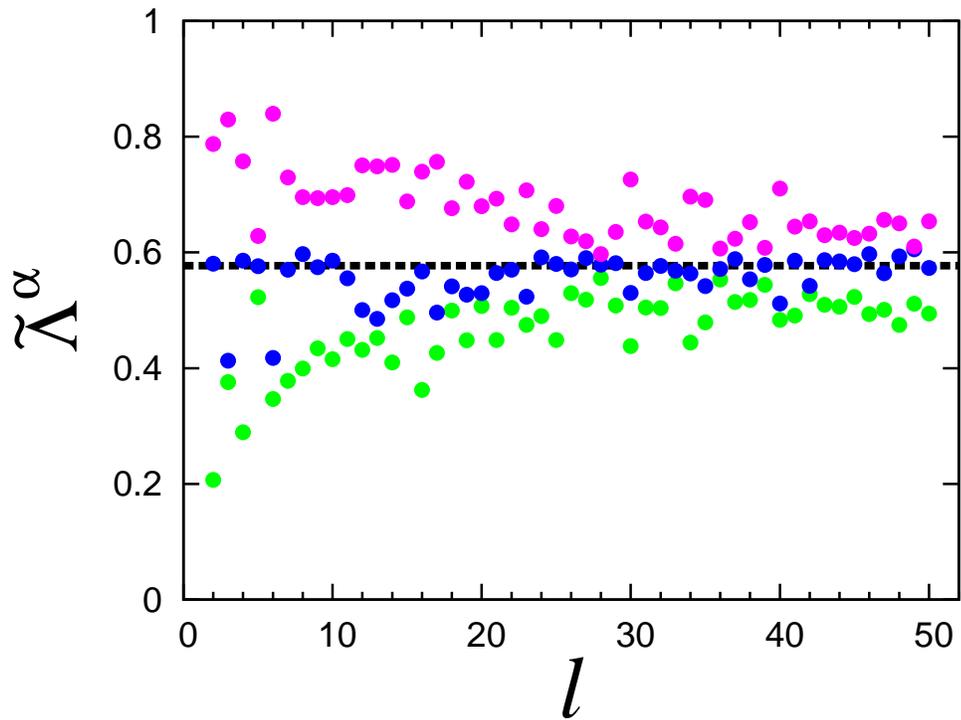}
\caption{ \small Normalized singular values versus $l$ for the ILC full sky data set. For each $l$ there are 3 values $\tilde \Lambda^{\a}$ normalized so the sum of their squares is unity.  The dashed line shows the isotropic prediction $\tilde \Lambda^{\a} =1/\sqrt{3}$.  }
\label{fig:ILCSingValsVSl.eps}
\end{center}
\end{figure}

\begin{figure}
\begin{center}
\includegraphics[width=4in,angle=-90]{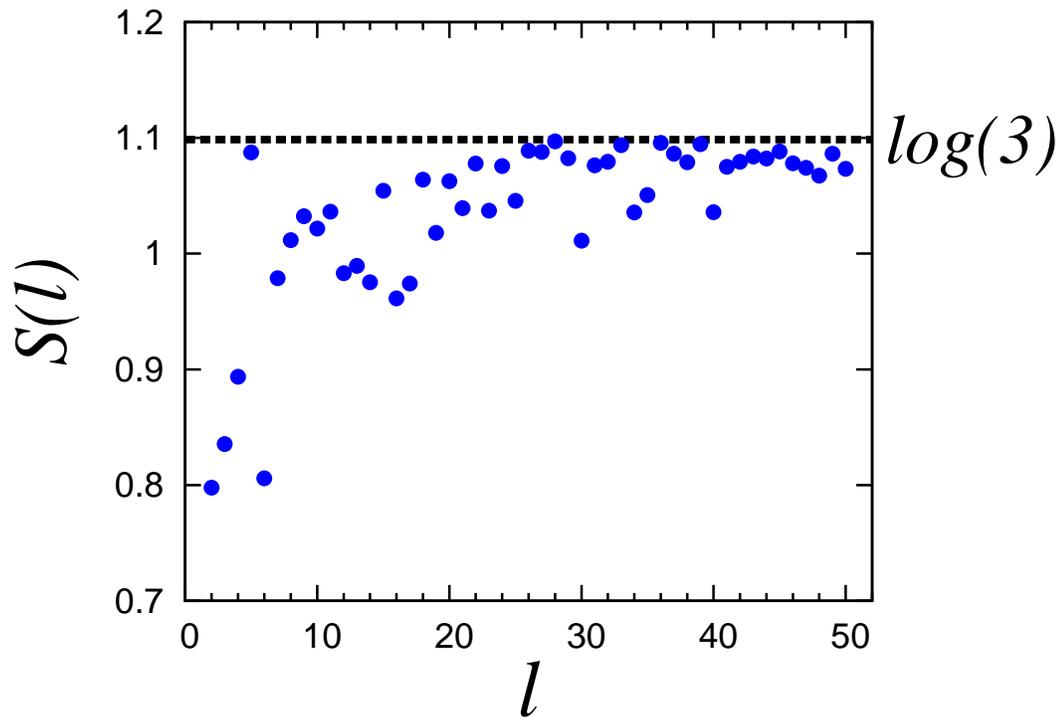}
\caption{ \small Power entropy $S(l)$ of the ILC full sky singular values versus mode number $l$. Low power entropy indicates orderly behavior not typical of isotropic data. The dashed line corresponds to the maximum values of the 
entropy $S=log(3)$.  }
\label{fig:ILCentropyVSl.eps}
\end{center}
\end{figure}

Fig. \ref{fig:ILCSingValsVSl.eps} shows the normalized eigenvalues $\tilde \Lambda^{\a}$, with a dashed line for the isotropic prediction $\Lambda_{\a} =C_{l}/\sqrt{3}$.  The normalization is $\sum_{\a} \,(\tilde \Lambda^{\a})^{2}=1$.  The power entropy $S_{full-sky}(l)$ mode by mode in $l$ for the ILC map is shown in Fig. \ref{fig:ILCentropyVSl.eps}. By Monte Carlo simulations we find that
the points $l$= 6, 16, 17, 30, 34, 40 appear to be statistically unlikely.

\begin{figure}[htp]
\centering
\includegraphics[width=4in,angle=-90]{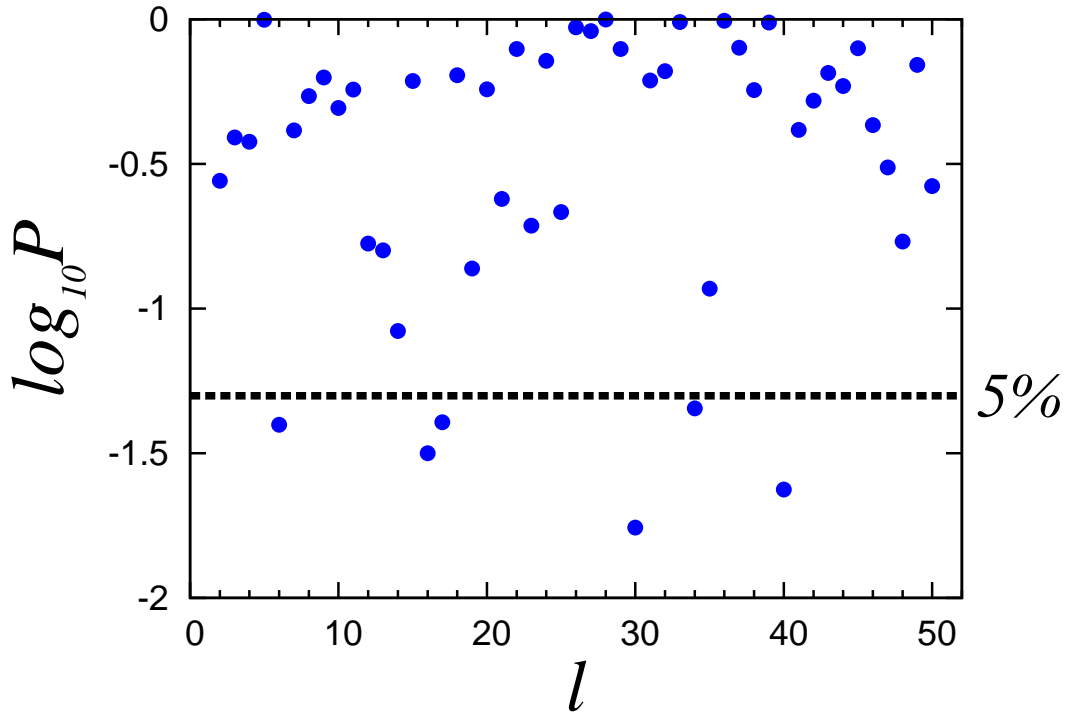}
\caption{ \small $log_{10}(P)$-values of the full-sky ILC map power entropy shown in Fig. \ref{fig:ILCentropyVSl.eps}. $P$-values are estimated from 10,000 realizations of random isotropic CMB maps. The dashed horizontal line shows $P=5\%$.  }
\label{fig:PowEntyPvals2-50.eps}
\end{figure}

We generated random $a_{lm}$ and derived the power entropy using 10,000 realizations of  CMB maps. In our procedure real $a_{l0}$ values were drawn from Gaussian distributions with zero mean and unit norm. Complex $a_{lm}$ for $m=1,..,l$ were created by adding real and $i$-times real numbers from the same distribution and dividing by $\sqrt{2}$.  As a consistency check, $P$-values were computed independently by different members of the group, both including the $C_{l}$ values, and omitting them, to verify the method and that the usual power statistic indeed drops out. Finally we computed $P$-values representing the frequency that power entropy in a random sample is less than that seen in the map.  In Fig. \ref{fig:PowEntyPvals2-50.eps} we show $P$-values of power entropy realized in the ILC maps versus mode number $l$.   
There are  6 unusually low power entropies of 5\% or smaller found at 
$l=$6, 16,  17, 30, 34 and 40, with $P$-values of 
0.040, 0.032, 0.041, 0.018, 0.045, 0.024 respectively.

It is interesting that the power entropy does not single out $l=2$ or 3 as particularly significant. The power entropy itself does not have any information about the {\it directional alignment} of eigenvectors, which is independent and will be studied separately. 

Note there are many {\it independent} instances of $P$-values less than the expected average of 1/2.  What is the probability to see so many small $P$-values? Since the $P$-values already take into account the distributions, this basic question has an analytic answer. 

There are 6 power entropies with $P$ values of $p=0.045$ or less.  The distribution to get $k$ independent $P$-values, found less than or equal to $p$, among $n$ independent trials is $f( \,k  \, | \, n,  \, p )=p^{k}(1-p)^{n-k} n!/(n-k)!k! $  This is the binomial distribution, which comes from counting $k$ instances of ``passes'' when considering ``pass-fail'' questions, while accounting for $n!/((n-k)!k!)$ independent ways to get the outcome ``pass.''  
The total number of independent trials with $l_{\rm max}=50$ are $n=48$.
The total probability of the data given random chance is\footnote{In discrete 
data analysis each particular possibility is enumerated, as in throwing dice, so we put no emphasis on the cumulative distribution. The cumulative binomial probability to see 6 or more cases is slightly larger, $f(k \geq 6  \, events \, | \, n=48, \, p=0.045)  \sim 0.020.$ }
\ba f( \,6 \, events \, | \, n=48, \,p= 0.045 ) =0.015 \ea  
This clearly shows that the signal of anisotropy is present over a much larger
$l$ range in the CMB data in comparison to what is commonly believed in
the literature ( Katz and Weeks 2004, Bielewicz
  {\it et al} 2004, Bielewicz {\it et al} 2005,
    Prunet {\it et al} 2005, Copi {\it et al} 2006,
      de Oliveira-Costa and Tegmark 2006, 
   Bernui {\it et al} 2006, Freeman {\it et al} 2006, Magueijo and  Sorkin 2007,
Bernui {\it et al} 2007,
 Copi {\it et al} 2007,  Helling {\it et al} 2007, Land and Magueijo 2007).

It is interesting to throw out unlikely events one by one. The  binomial probability to find 6, 5, 4... small $P$-values sorted from  0.045 to smaller levels is $1.5 \times 10^{-2},\, 3.1 \times 10^{-2} , \, 7.5  \times 10^ {-2}, $... respectively. We also conducted a  Monte Carlo simulation to verify the distribution is as stated.

A common criterion to test hypotheses poses confidence level break-points of $5\%$. The distribution of the invariant power entropy is not consistent with the isotropic model. 
 
\subsection{Alignments} 

The alignment of eigenvectors in the \emph{ILC Map} at low $l$ may be a signal of a fundamental anisotropy or could also be caused by galactic foreground contamination.  
The history of alignment with Virgo suggests we compare the principal axes of the power tensors $l$-by-$l$ with the quadrupole axis. 

\begin{figure}
\begin{center}
\includegraphics[width=4in]{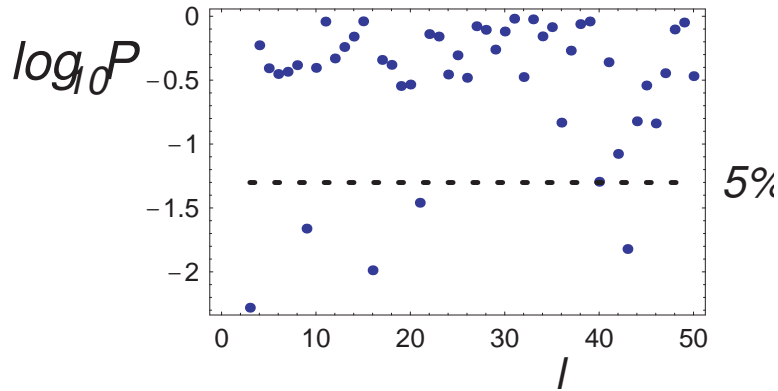}
\caption{$log_{10}(P)$-values of largest power eigenvectors to align with the quadrupole versus $l$.  The dashed line shows the $5\%$ level of significance. There are 6 points at the level of $P \lesssim 5\%$ or less. }
\label{fig:BigEvalPvals.eps}
\end{center}
\end{figure}

Fig. \ref{fig:BigEvalPvals.eps} shows $log_{10}(P)$-values in an  isotropic distribution for power principal eigenvectors to align with  the quadrupole.  There are 6 axes aligned with a significance close  to $5\%$ level or less; they are multipoles with $l=$ 3, 9, 16, 21,  40, 43. The binomial probability of seeing 6 such events is 2.3\%. 
 Hence the alignment with the quadrupole
 is seen over a relatively large range of $l$ values.
If coincidences are thrown out one by one, the estimated probability  to find (6, 5, 4...) small $P$-values seen in the data is $2.3 \times  10^{-2} , \, 1.7\times 10^{-2}, \, 1.4\times 10^{-2}$.... 
The  distribution of the principal axes for many different $l$ are not at  consistent with the isotropic proposal.  The Cartesian components of  the well-aligned principal axes are listed in Table \ref{tab:Xtable}.

\begin{table}
  \centering 
  \begin{tabular}{cc}
\hline
\hline
$l$  & $ axis \, \hat n \, (\,\, x, \, y, \, z \, \,)$ \\
\hline
\hline
2&$(\,-0.209,\,-0.302,0.930\,)$\\
3&$(\,-0.251,\,-0.385,\,0.888\,)$\\
9&$(\,-0.224,\,-0.0974,\,0.970\,)$\\
16&$(\,-0.157,\,-0.175,\,0.972\,)$\\
21&$(\,-0.139,\,-0.536,\,0.832\,)$\\
40&$(\,-0.0928,0.211,0.973\,)$\\
43&$(\,-0.217,\,-0.132,\,0.967\,)$\\
\hline
1&$(\,-0.0616,\,-0.664,\,0.745\,)$\\
$\hat n_{X} $ & $(\, -0.435,  \,0.134,  \,0.890 \,) $ \\
\end{tabular}
  \caption{Galactic Cartesian coordinates of principle axes (``large eigenvectors'') $\hat n$ that are extraordinarily aligned.  Criteria for selection are $P-values \lesssim 5\%$ relative to the $l=2$ principal axis. The $l=1$ or dipole axis is included for completeness.  The axis $\hat n_{X} $ is the principal eigenvector of the directional entropy of the set $2 \leq l \leq 50$. }
 
  \label{tab:Xtable}
\end{table}

It may appear arbitrary to compare the axes to the $l=2$ case. It is motivated by the literature and our previous history of work, discussed in the Introduction. For completeness $P-values$ among all the independent pairs are listed in Table \ref{tab:Ptable}.  It shows that choosing the $l=2$ case causes no special bias, because once the axes are well-aligned any one of them could be used for comparison.  Correlation with the dipole, which is not strictly part of the ILC map, is also included for reference in Table \ref{tab:Xtable}.  The statistical significance against isotropy is high whether or not one includes the dipole. 

\begin{table}
  \centering   
  \begin{tabular}{cccccccc}  
\hline 
\hline 
& $l=$ 2 & 3 &9  & 16  & 21 & 40  &  43 \cr 
\hline 
$l' =2$ &  .  & .  & .  & .  & .  & .  & .  \cr
$\,3$ & 0.005 & . & . & . & . & . & . \cr
\,9 &  0.022 &  0.045 & . & . & . & . & . \cr 
\,16 &    0.010 & 0.030 & 0.005 & . & . & . & . \cr 
\,21 &  0.035 & 0.019 & 0.109 & 0.075 & . & . & . \cr
\,40 &  0.051 & 0.078 & 0.057 & 0.032 &    0.090 & . & . \cr
\,43 &  0.015 & 0.036 & 0.0006  & 0.003 & 0.094 & 0.051 & . \cr 
\hline 
\,1 &   0.094 &     0.067        &    0.199         &    0.150     &    0.015        &  0.141       &   0.178        \cr
 \end{tabular}
  \caption{ \small$P-$ values of coincidence between independent pairs of principal axes labeled by $l, \, l'$ shown in Table \ref{tab:Xtable}.  Correlation with the $l'=1$ axis, which is not strictly part of the ILC map, is included for completeness.  }
  \label{tab:Ptable}  
\end{table}


%

\subsubsection{Alignment Entropy} 

In performing this research we first tested for clustering using the directional entropy $S_{X}$ of matrix $X$. Comparison with a Monte Carlo simulation using the same range $2\leq l \leq 50$ does not show anything extraordinary, yielding a $P$-value of $69.0$ \%. The directional entropy $S_{X}$ did not signal significant clustering.  Yet the directional entropy is rather insensitive statistic which misses the pattern in Table \ref{tab:Ptable}.   

Curiously, however, the principal axis of $X$ is \ba \hat n_{X} = (\,-0.435,  \,0.134,  \,0.890  \,). \nn \ea   
The angular galactic coordinates $(\, l= 162.88, \, b= 62.91\,)$ are clearly not in the galactic plane, but well aligned with Virgo.   Recall that the principal vectors of the quadrupole and octupole in the \emph{ILC Map} have galactic Cartesian components 
\begin{eqnarray}
\hat n (l=2)=(\,-0.209,\, -0.302,\, 0.930\, ) \nonumber \\
\hat n (l=3)=(\, -0.251,\, -0.386,\, 0.888 \,)
\end{eqnarray}
respectively. These two axes are very closely aligned with one another, as so much noted.  It is remarkable that they are aligned with the principal axis of $X$ coming from the whole ensemble. 

\begin{figure}[htp]
\centering
\includegraphics[scale=0.35,angle=-90]{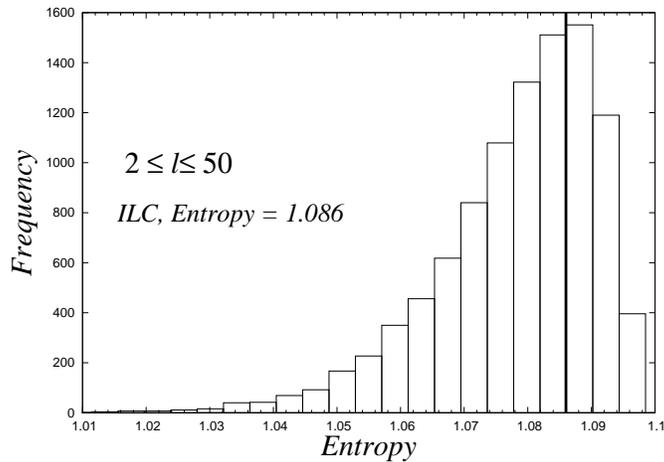}
\caption{ \small Directional entropy histograms of $S_{X}$
for the multipole range 
$2 \leq l \leq 50$. The value of  
directional entropy of the ILC map are also shown. Histograms were generated using
10,000 randomly generated CMB data sets.
}
\label{fig:aniso}
\end{figure}

\begin{figure}[htp]
\centering
\includegraphics[scale=0.42,angle=-270]{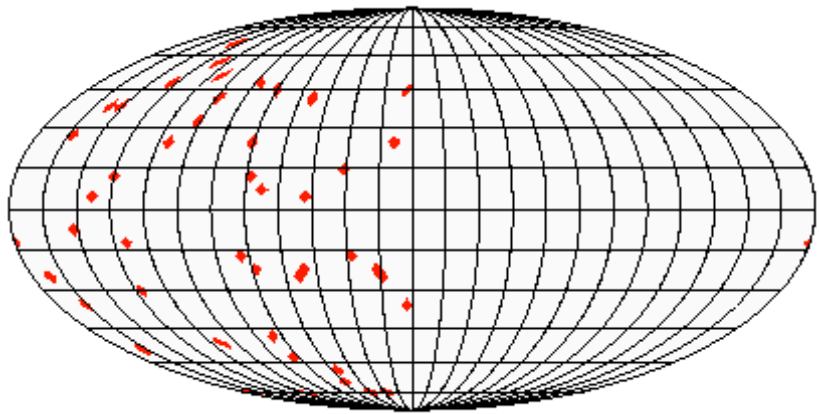}
\caption{ \small Mollweide projection plot in galactic coordinates of principal axes for the \emph{ILC map} over the range $2 \leq l \leq 50$.  No clustering is visible because the cluster of axes near the galactic pole is distorted.}
\label{fig:mollview}
\end{figure}

\begin{figure}
\begin{center}
\includegraphics[width=4.5in]{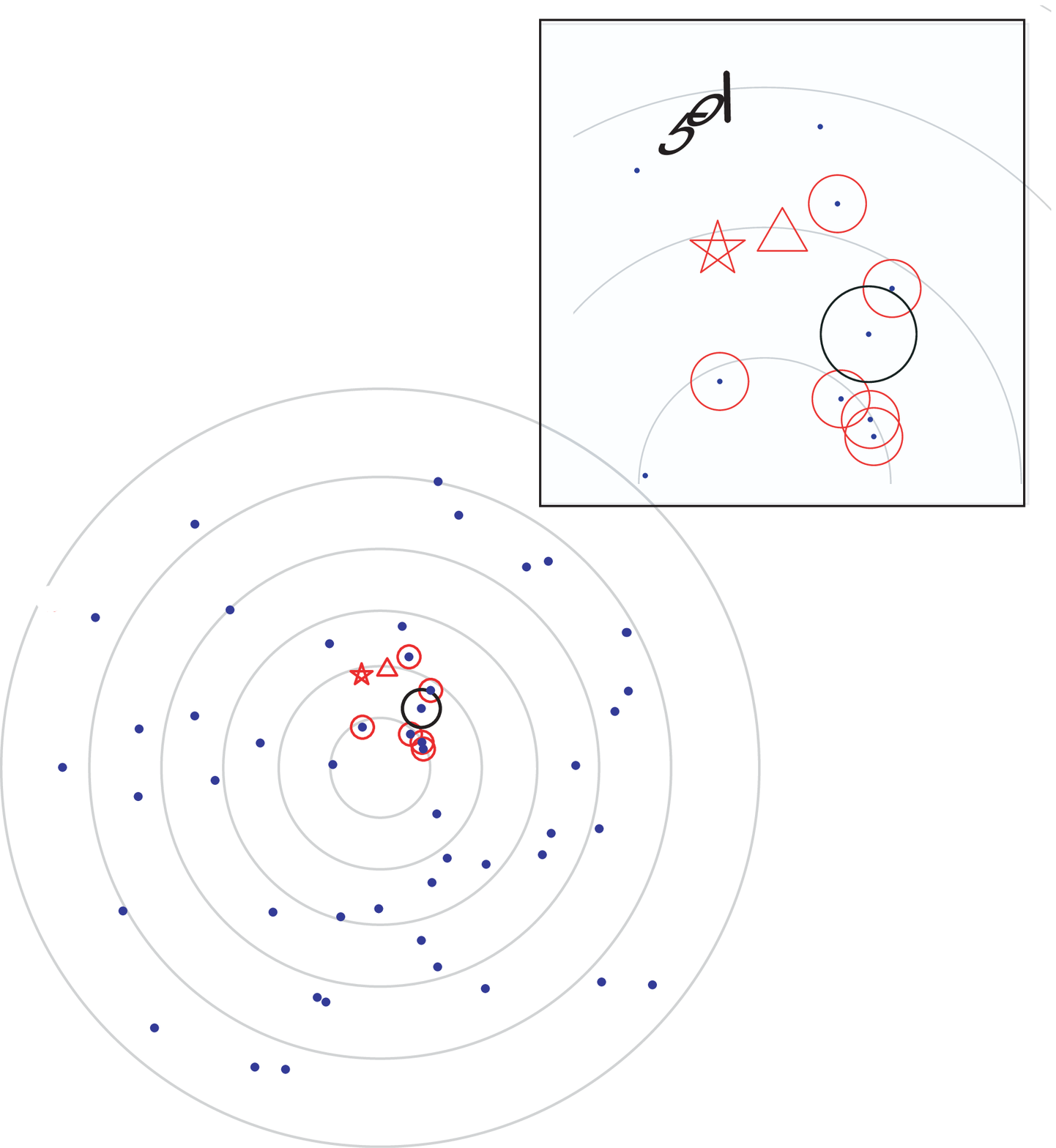}
\caption{Clustering of principal axes visible in stereographic projection from the North galactic pole. Highly correlated axes (Table \ref{tab:Xtable} are circled, with the quadrupole axis as reference point (large circle).  Also shown are the dipole (diamond), the radio polarization offsets alignment axis (star), and the QSO optical polarization alignment axis (triangle), as discussed in the text.  The inset box shows an enlargement and $5^{o}$ separation scale in the cluster region. }
\label{fig:stereo}
\end{center}
\end{figure}

Our first studies also used a Mollweide projection to examine visually for clustering of axes (Fig. \ref{fig:mollview}). Since we are plotting an axis, and not a vector, we plot points only on one-half of the sphere. The \emph{ILC map} does not visually show any striking
clustering among different eigenvectors in Mollweide projection.

However the Mollweide projection seriously distorts the region near the galactic poles.  Fig. \ref{fig:stereo} shows a stereographic projection, a method that is free of distortion near the poles. Here the clustering of axes is clearly visible, explaining that the Mollweide projection has sent all the interesting points out of view.  

\subsubsection{The Region $1 \leq l \leq 11$ }

The method of Copi {\it et al} (2007) finds highly significant signals for 
the region $1 \leq l \leq 11$. 
It is interesting to evaluate our results for this range.

There are {\it four}  well-aligned principal vectors at $l=$ 1, \, 2, \, 3,  9.  Removing one as trivial, the probability to find 3 $p$-values of less than 5\% probability in a random sample in 11 trials is about $1\%$.   {\it One} unusually low power entropy value exists in the range at $l=6$.  

\begin{figure}
\begin{center}
\includegraphics[width=4.0in,angle=-90]{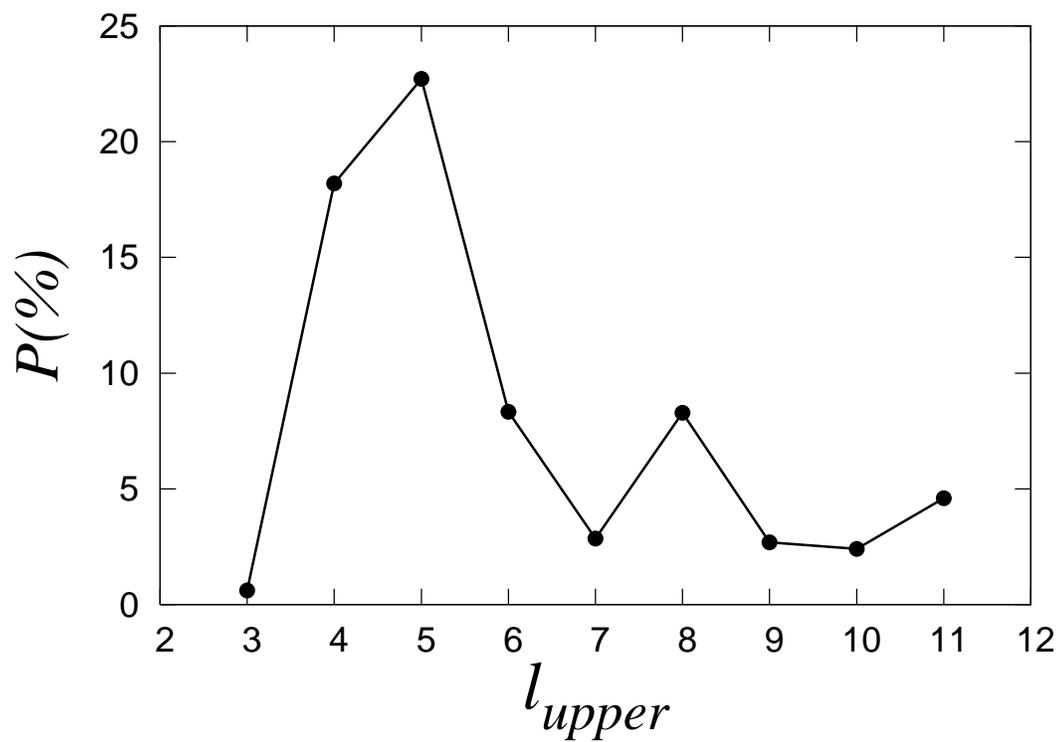}
\caption{The P-values (in \%) of the full-sky ILC map of the directional 
entropy $S_X$ in the range $2\le l \le l_{\rm upper}$. }
\label{fig:pvalues2-11}
\end{center}
\end{figure}

What about the directional entropy ?  When $S_{X}$ is evaluated for the range $2\le l\le 11$, random samples show $S_{X}$ equal or lower than the \emph{ILC Map} only 440 times out of 10,000, a $P$-value of $4.4\%$. This is another indication of anisotropy. The P-values for different upper limits $l_{\rm upper}$ 
are shown in Fig. \ref{fig:pvalues2-11}. We see that most of the P-values
in this range are smaller than 5 \%.
Finally the principal axis of $X$ is \ba \hat n_{X}(2 \leq l \leq 11)= (\,  -0.531, \, 0.117, \, 0.839 \,). \nn \ea  It is well-aligned with Virgo.  The signals from $1 \leq l \leq 11$ evidently come from the same source as the signals from the entire range $2 \leq l \leq 50$.

\subsubsection{The $Y$ Matrix} 

\begin{figure}
\begin{center}
\includegraphics[width=5in]{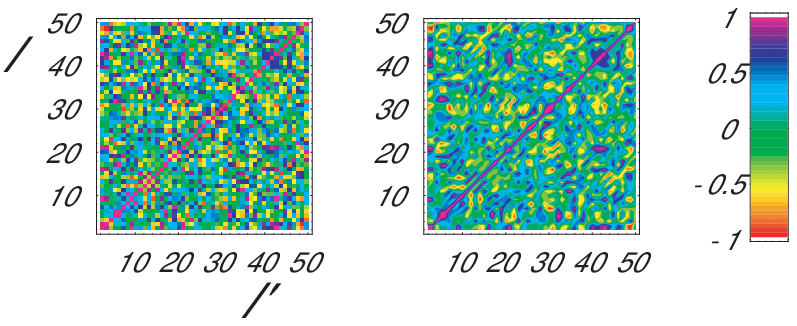}
\caption{Contour maps of the correlation $Y(l, \, l')$ of traceless tensor power between different angular momentum indices $l$. Left: a simple density plot.  Right: Contours interpolated with trends more readily visible. A scale (color online) shows the color code of correlation values.  Features confirm the other studies of unusual correlations in supposedly uncorrelated random CMB data. }
\label{fig:elcor2-50.eps}
\end{center}
\end{figure}

The $Y$-matrix gives another method to examine visually the angular correlations. 
Fig. \ref{fig:elcor2-50.eps} shows contour maps (color online) of the $Y(l, \,l')$ correlation.  The correlation in ideal uncorrelated data should show a diagonal line up to fluctuations.  Features of Fig. \ref{fig:elcor2-50.eps} confirm the other studies of unusual correlations in nominally uncorrelated random data. This study complements the study with matrix $X$, which uses principal axes and does not use the information contained in the singular values of $\psi$.

\subsection{Foreground Contaminated Maps}

Here we study effects of simulated foreground contamination on the power tensor and entropy statistics. 

\begin{figure}[htp]
\centering
\includegraphics[scale=0.8,angle=0]{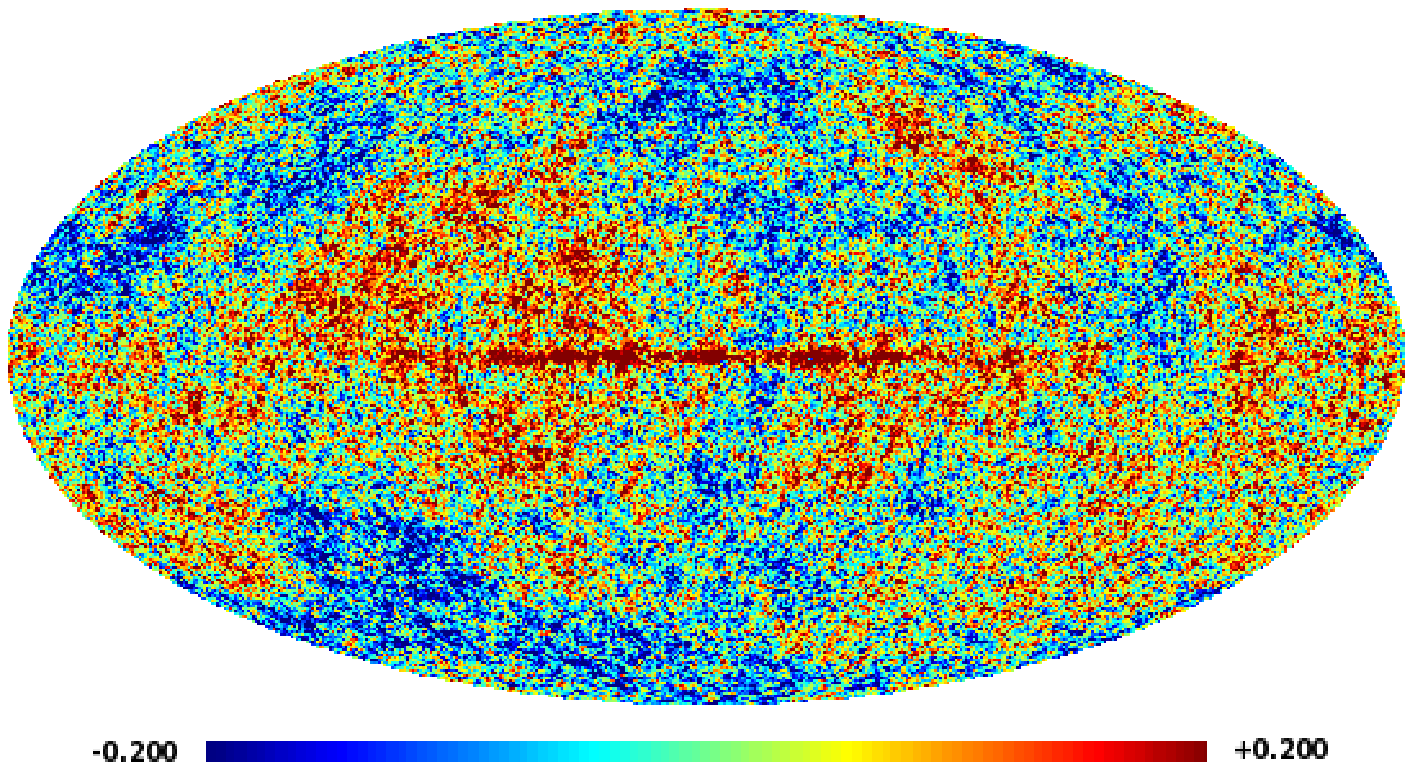}
\includegraphics[scale=1.48,angle=90]{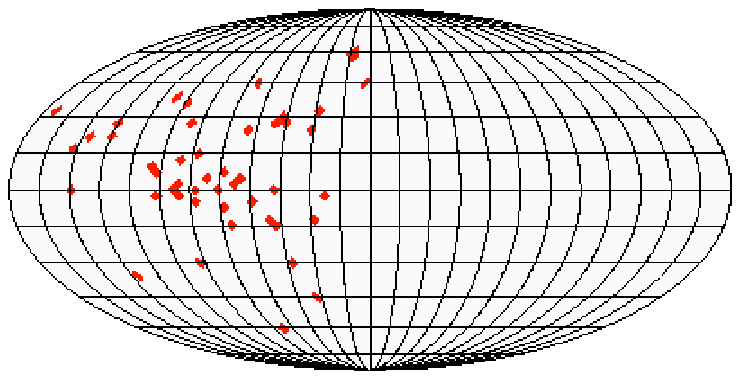}
\caption{ \small Top panel: Simulated CMB +  ($2\%$) synchrotron contaminated foreground map described in the text. The simulated foreground contamination is $2\%$ of the actual foreground strength. Bottom panel: Principal axes for $2 \leq l \leq 50$ made from the map of the top panel show alignment in the plane of the galaxy.  Mollweide projection in galactic coordinates. 
}
\label{fig:fgcmb}
\end{figure}

A contour plot shows the input of the simulation in Fig.\ref{fig:fgcmb}, top panel.  
The map is made by adding a random CMB map with a simulated synchrotron 
foreground map (Giardino {\it et al} 2002) generated at 23 GHz frequency and 
normalized to 
$2\%$ of the actual foreground strength. Contamination is dominantly in the 
galactic plane using the procedure of . 

Our procedure readily senses the contamination.  Fig.\ref{fig:fgcmb}, bottom panel, shows the spatial distribution of largest eigenvectors $\tilde e_{l}$ for the range of multipole  moments $2\leq l \leq 50$. The vectors lie dominantly in the galactic plane, providing a clear signal of alignment caused by galactic contamination. 

Calculation of the entropy of the largest eigenvectors $S_{X}(l_{max}=50)=0.995$ also
shows significant anisotropy. The $P$ value for this to occur in a random isotropic sky is $P=0.01$ \%.  The principal axis, or largest eigenvector of $X$ is $\hat n= (-0.233, \, -0.963,\, -0.132)$, which nicely corresponds to $(l=76.4, \, b=7.56)$, lying close to the galactic plane.

These studies show that our procedure detects galactic contamination in a simple and statistically reliable way. They also indicate that the Virgo alignment observed in the preferred directions of the ILC data cannot be attributed to in-plane galactic contamination. 

\section{Masked Sky Studies}

As mentioned in the introduction, our use of linear transformations allows the effects of window functions of masks to be handled by conventional and  transparent means.
In this Section we explore the effects of masks applied to different hemispheres of the galaxy.  Although masks distort power spectra, using symmetrical masking creates no biases. 

\begin{figure}[htp]
\centering
\includegraphics[scale=0.95,angle=0]{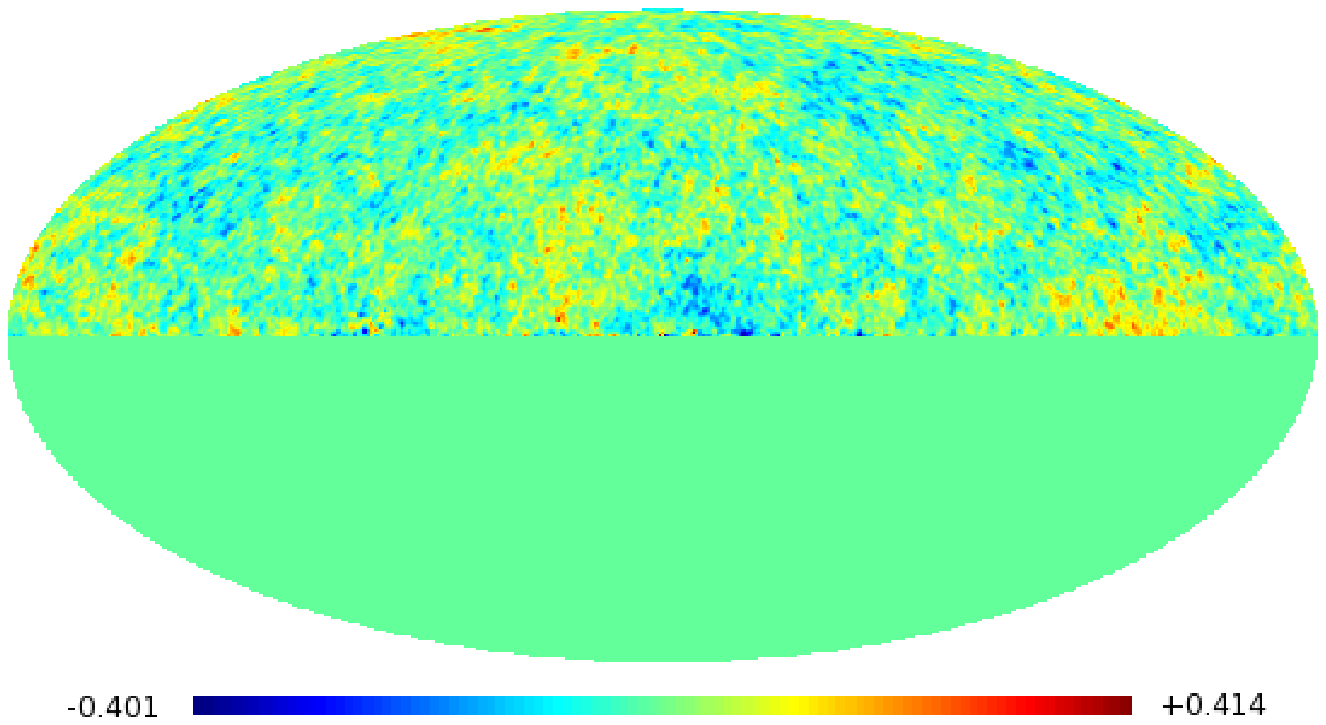}
\includegraphics[scale=0.95,angle=0]{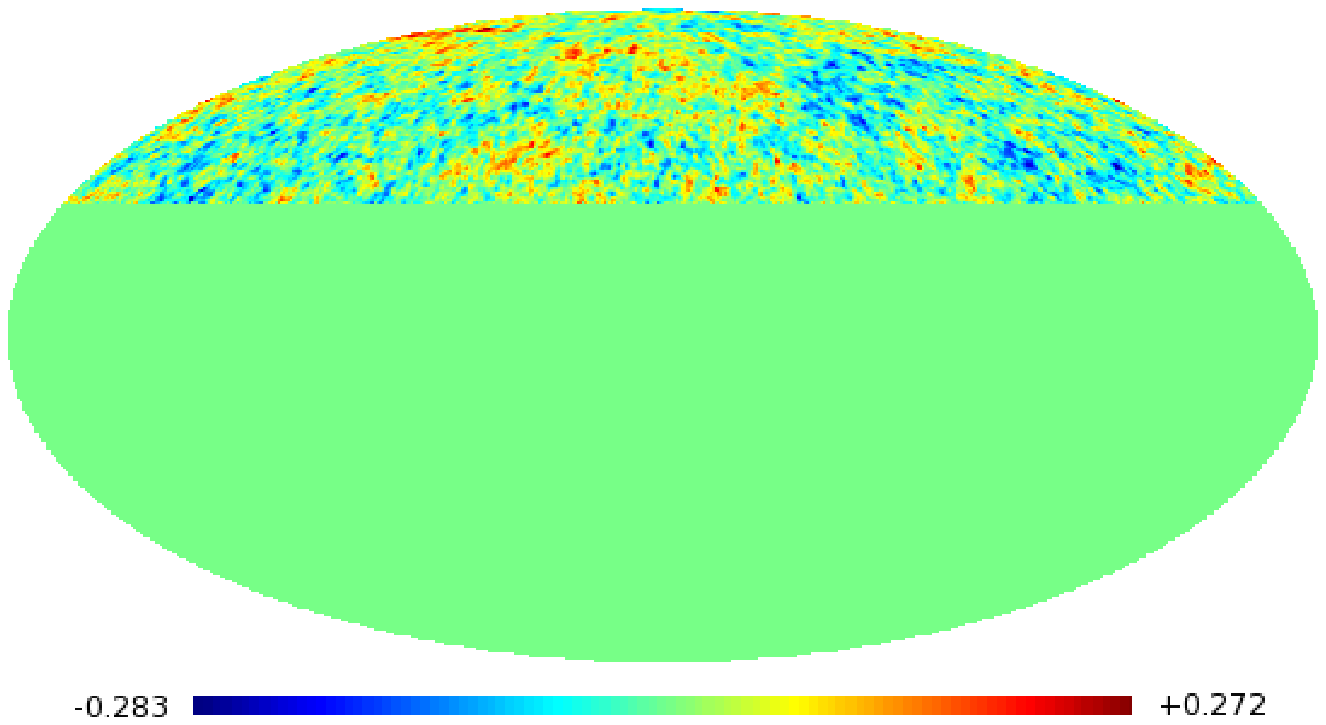}
\caption{ \small The WMAP-ILC map after applying masks $M1(\hat n)$ and
$CM1(\hat n) $.}
\label{fig:masks}
\end{figure}

We use four different masks. The Northern and Southern hemispheres  in galactic coordinates are retained in \emph{Mask 1} ($M1(\hat n)$) and 
\emph{Mask 2} ($M2(\hat n)$), respectively.  \emph{Cone Mask 1} 
($CM1(\hat n)$) and \emph{Cone Mask 2} ($CM2(\hat n)$) respectively retain the polar angle regions $0 \leq \theta \leq 30^\circ$ and 
$150^\circ \leq \theta \leq  180^\circ$. These
are typical values chosen to minimize galactic contamination.  We show the ILC Map cut by \emph{M1} and \emph{CM1} in Fig. \ref{fig:masks}.
 
\subsection{Isotropic Maps Estimated from Masked Maps} 

We explore the power estimated for a full-sky map, as commonly developed from the power spectrum obtained from unmasked regions.  A window function $W(\hat n)$ is defined so that \ba \Delta T^{mask}(\hat n) = W(\hat n)\Delta T(\hat n) . \nn \ea This relation is linear and invertible, and already in its diagonal form: \ba \Delta T(\hat n) =\Delta T^{mask}(\hat n)/W(\hat n) \label{win} \ea  Inversion is unstable (``ill-conditioned'') for small $W(\hat n)$, the region where information does not exist, a fact that naturally limits to information that can reliably be obtained. 

We used the method of Hivon {\it et al}, as we briefly review.  Let $a_{lm}(mask)$ be the amplitude obtained from $\Delta T^{mask}(\hat n)$.  The transformation of Eq. \ref{win} can be carried out in angular momentum coordinates, with the general form \ba a_{lm}(mask) = {\cal W}_{mm'}^{ll'}a_{l'm'}, \nn \\ \ket{a(l)(mask) }={\cal W}^{ll'}\ket{a(l')}, \label{power} \ea where ${\cal W}$ is the transformed kernel. In general all components of $a_{lm}(mask)$, and not just the power spectrum are needed.  The power spectrum is related by \ba \bracket{a(l)(mask)}{a(l)(mask)} = \EV{a(l'')}{ \sum_{l} \, { \cal W}^{l''l*} {\cal W}^{ll'}} {a(l')}. \label{pow2} \ea  An estimate of the power spectrum from inverting Eq. \ref{pow2} will be linear but generally couple all multipoles in complicated ways. 

To construct an inversion Hivon {\it et al} make the assumption of isotropy.  In Eq. \ref{pow2} replace \ba\bracket{m}{a(l)} \bracket{a(l'')}{m''} \ra C_{l}\delta^{ll''}\delta_{mm''}. \ea Assuming this applies to ensemble averages, the linear relation between power measures is:


%

%

\begin{eqnarray}
< C_{l}(mask)> = \sum_{l^\prime}  \,  M_{ll^\prime}<C_{l^\prime}>,
\label{mode_coupling}
\end{eqnarray}
where $M_{l_1l_2}$ is computed with Clebsch ($3j$) coefficients
\begin{equation}
M_{l_1l_2}= \frac{2l_2+1}{4 \pi}\sum_{l_3=|l_1-l_2|}^{l_3=l_1+l_2}(2l_3+1)W_{l_3}
\left(
\begin{array}{ccc}
l_1 & l_2 & l_3 \\
0   &  0  &  0
\end{array}
\right)^2
\end{equation} Here $W_{l}$ is the power spectrum of the mask function. Beam and pixel factors are readily incorporated in Eq. \ref{mode_coupling}, while for our purposes they can be absorbed into the symbols.   Since Eq. \ref{mode_coupling} is linear, the kernel $M$ can be inverted to solve for estimated $C_{l}$ of an ideal full sky in terms of the observable power $ C_{l}(mask)$.   

In general inversion remains ill-conditioned, and care must be used in interpretation.  Hivon {\it et al} choose to work with binned power spectra defined by a running average:  \ba  \tilde C_{b}= \sum_{l} \, P_{bl} C_{l}(mask). \  \label{binned} \ea The ``binning operator'' $P_{bl} $ we use averages 3 multipoles, with $$ \tilde C_{b}= \sum_{l = l^b_{min}}^{l^b_{max}} \,\frac{l(l+1)}{2\pi} C_{l}(mask)/3$$. 
Binning may ameliorate correlations caused by windowing, while generally creating its own correlations between different $\tilde C_{b}$. The estimated full sky power spectrum ${\mathcal{C}_{b^\prime}}$ is \ba {\mathcal{C}_{b^\prime}} =  \sum_{b'} \, K^{-1}_{b^\prime b}\tilde C_{b}; \nn \\ K_{bb^\prime}=(P\cdot M\cdot Q)_{bb^\prime} \nn \ea
The ''Inverse binning operator" $Q_{lb}$ is given by $Q_{lb} = \frac{2\pi}{l(l+1)}$ and is nonzero only if $2\le l^b_{min} \le l \le l^b_{max}$. This summarizes sophisticated data processing procedures applied to the BOOMERANG experiment and many others.  
 
\begin{figure}[htp]
 \centering
\includegraphics[scale=0.4,angle=-90]{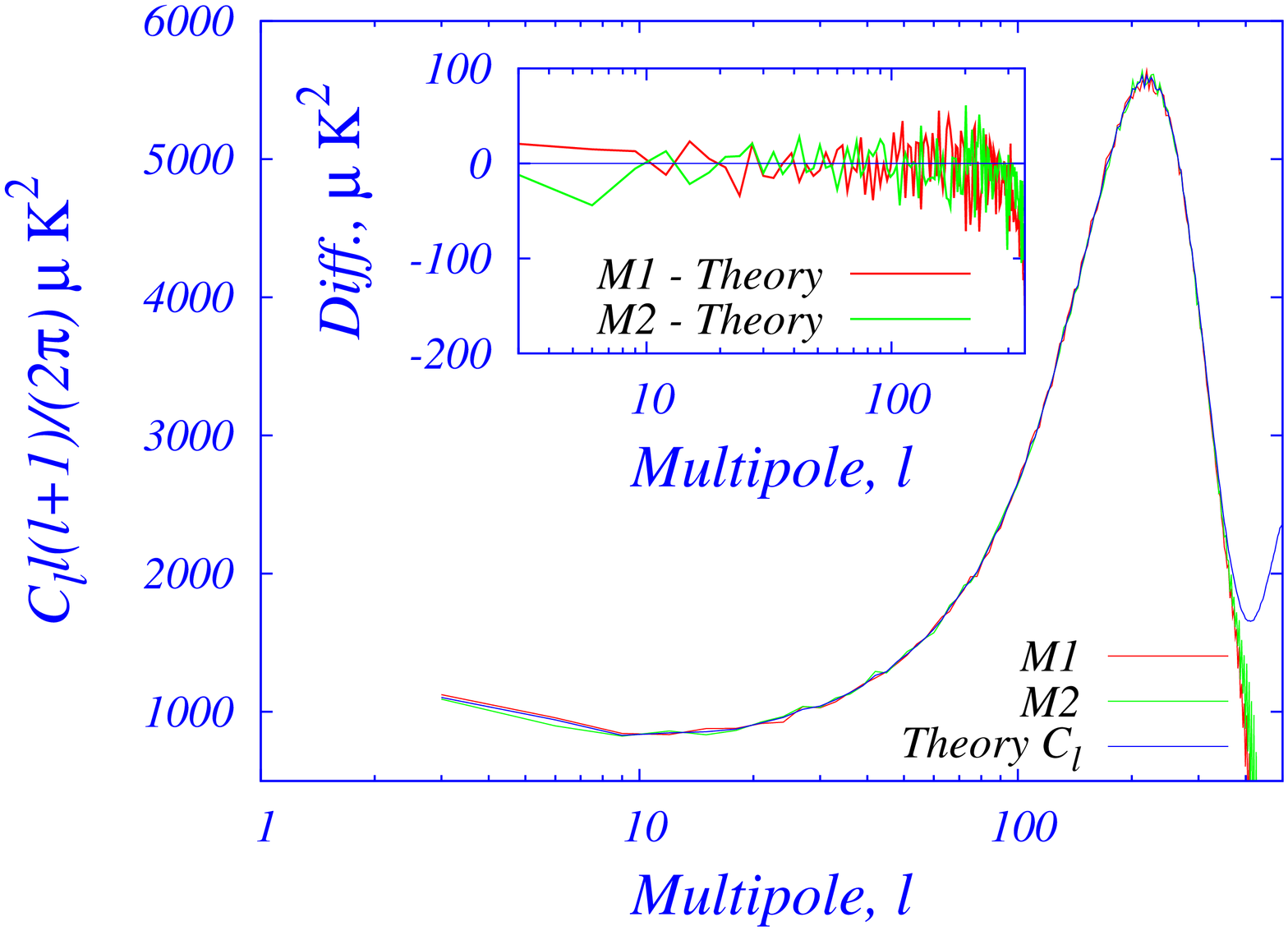}
\includegraphics[scale=0.4,angle=-90]{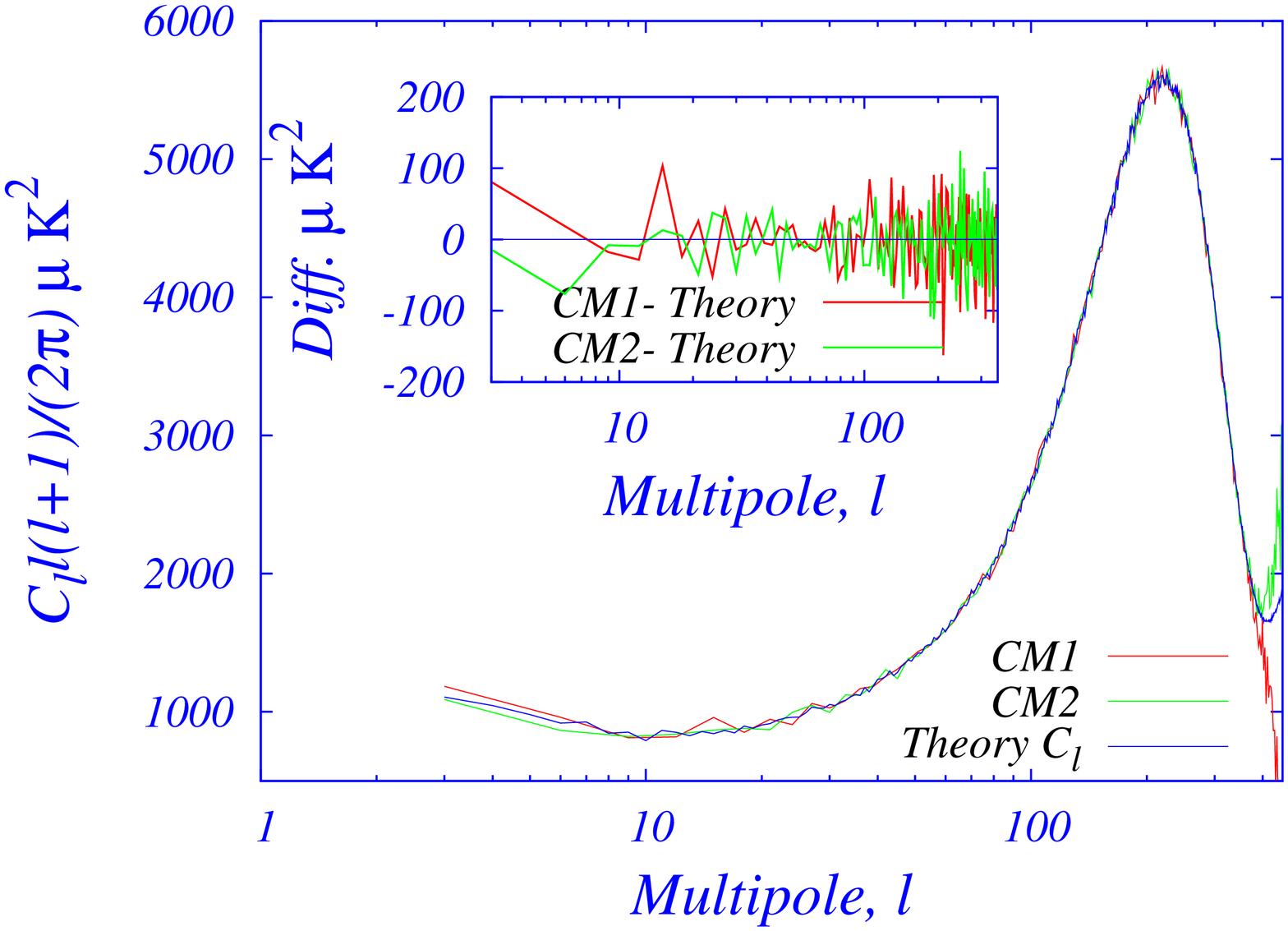}
\caption{ \small Comparison of estimated full sky power spectrum with input 
power spectrum simulation using only masked regions.  Blue (red) online denote the Northern and Southern Hemisphere cases respectively. Top panel: {\it Mask 1} and {\it Mask 2}.  Bottom panel: {\it Cone Mask 1} and {\it Cone Mask 2}.  The spectra match the average of the spectra generating them (solid black line, ``Theory'') nearly to the upper limits of power developed in the maps, $l \sim 300$.  The inset shows the difference {\it Diff}. }

\label{fig:simulation}
\end{figure}

We tested our code for full sky power spectrum estimation using Monte-Carlo simulation. We generated 100 random CMB maps, masked them by 
each of the four masks, and obtained partial sky power spectra.  We binned the average masked sky power coefficients to decrease dimensions from an original size of $1024 \times 1024$ to a reduced size of $341 \times 341$. Finally we produced an estimated full sky power. Figure \ref{fig:simulation} shows the results of the simulations. The estimated full sky power spectrum matches quite well with the average of the input full sky binned power spectra for each of the four masks. 

We note that the binning and isotropic assumptions of the full sky estimation method naturally dilute statistical measures.  We did not pursued the systematic errors.  We simply applied our procedures and studied the outcomes. 

\subsection{Features of Estimated Full Sky Spectra } 

\begin{figure}[htp]
\centering
\includegraphics[scale=0.35,angle=-90]{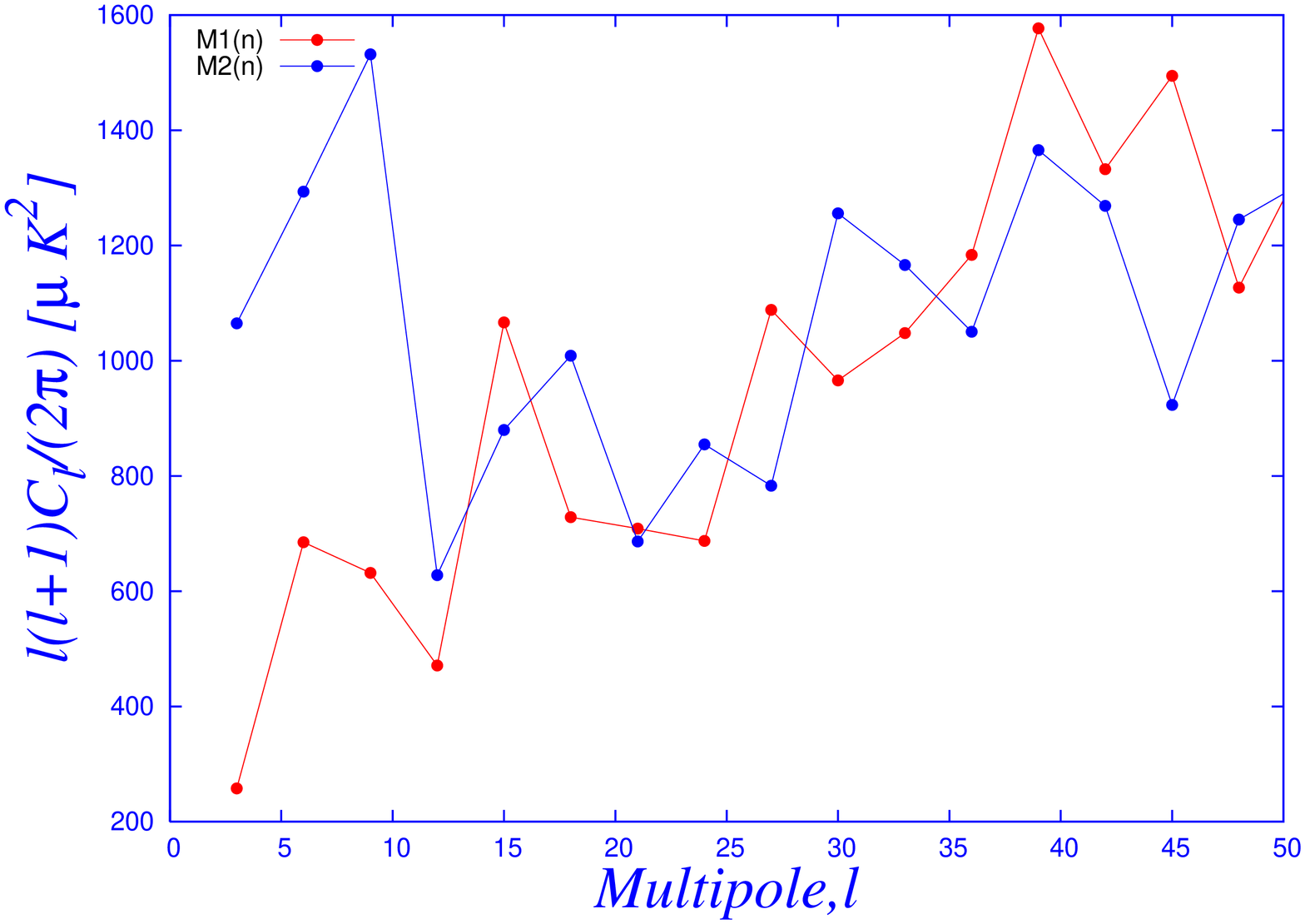}
\includegraphics[scale=0.35,angle=-90]{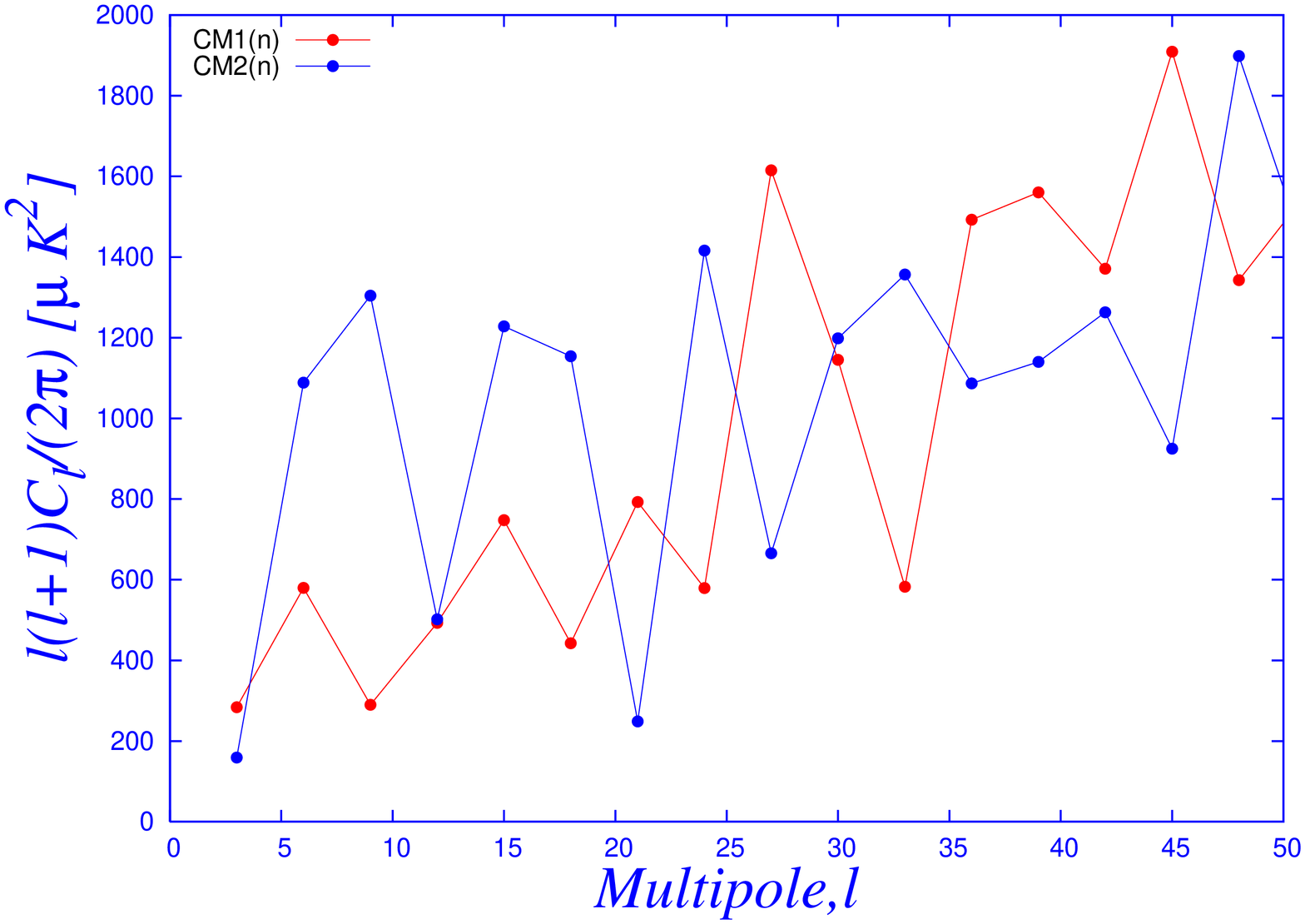}
\caption{ \small Estimated full sky binned power 
spectra, Northern and Southern galactic hemisphere, using 
$M1(\hat n)$ and $M2(\hat n)$(upper panel) and $CM1(\hat n)$ and $CM2(\hat n)$ (lower panel).}

\label{fig:result}
\end{figure}

In Figure \ref{fig:result} we show the 
{\it estimated} full sky power spectra obtained from ILC using the four masks
$M1(\hat n)$, $M2(\hat n)$ and $CM1(\hat n)$, $CM2(\hat n)$.  
We observe a definite asymmetry in the estimated full-sky power spectrum between the Northern and Southern hemisphere in several multipole ranges. 


It is interesting that $M1(\hat n)$, $M2(\hat n)$ and $CM1(\hat n)$, $CM2(\hat n)$ find similar excess in the same multipole ranges. This shows that the power anisotropy observed cannot be explained away by contamination of  the galactic plane, but also exists at high galactic latitude.  To quantify the power differential we sum the difference in power over all bins.  The probability of randomly getting an equal or larger difference is less than 
1 \% for the range $2\leq l \leq 11$ and about 9 \% for the range 
$2 \leq l \leq 50$. These figures are given retrospectively after scanning the figures, and one may feel they represent a search.  Given the unknown systematic errors we will not bother estimating significance. It is nevertheless interesting that 
a strictly power-based search, which includes averaging features that tend to wash out signals, can still potentially detect anisotropy. 
 
\begin{figure}
\begin{center}
\includegraphics[scale=0.3,angle=-90]{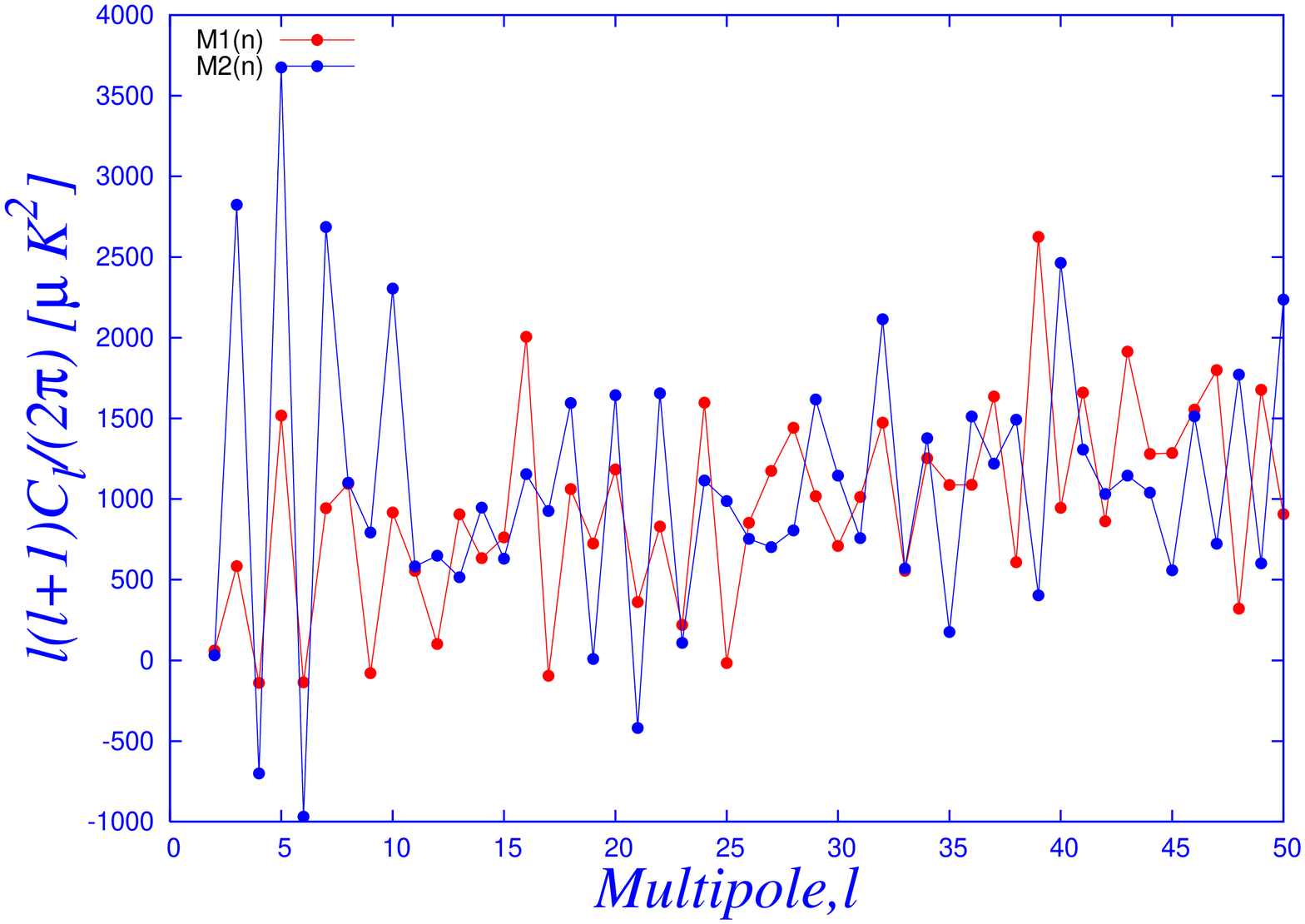}
\caption{The estimated full sky power map constructed from Mask 1 {\it without} going through the running average (``binning'') procedure that combines different $C_{l}$.  Note that negative power is sometimes predicted.  }
\label{fig:ilcUnbinnedHivonWay.ps}
\end{center}
\end{figure}

Just for interest we also include the estimated full sky ILC power from an {\it unbinned} inversion (Fig. \ref{fig:ilcUnbinnedHivonWay.ps}).  There are substantial differences from the binned evaluation. For example, negative power occurs in some cases. Since the unbinned procedure retains {\it more} information about the inversion than binning, it is a symptom of the instability of inversion, and not a mathematical error.  Use of binning helps to cure the problem for an ensemble of many maps, but it weakens the relation of any particular map to any particular data.  Exploring such features gives an idea of the reliability of trying to improve instabilities of inversion by binning. 

\section{Summary and Conclusions}

\begin{figure}
\begin{center}
\includegraphics[width=4in,height=2.5in]{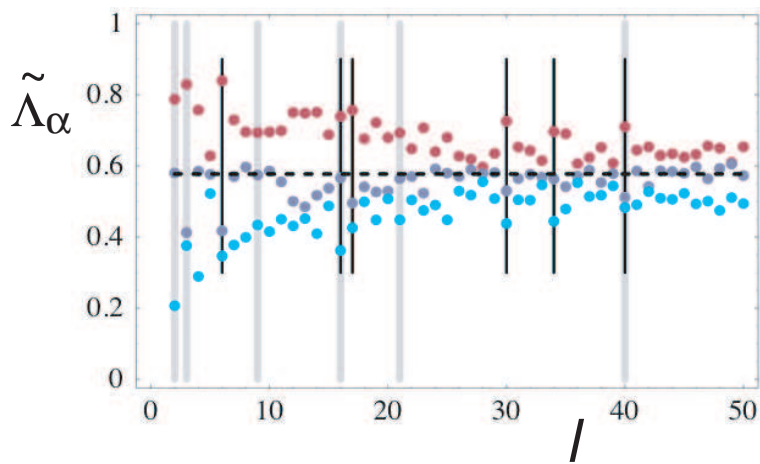}
\caption{ \small Normalized singular values versus $l$ with exceptional cases highlighted. Thin short lines denote $P$-values of entropy $S_{power} \lesssim 5\% $.  Thick gray lines denote principal axes aligned with the quadrupole (Virgo) axis with $P \lesssim 5\% $. There are 12 independent cases of 5\% or lower significance, which is unlikely in isotropic data at significance exceeding $99.9 \%$. Significance is increased when the dipole is included. }
\label{fig:SinValPVal2.eps}
\end{center}
\end{figure}

We have developed several new symmetry-based methods to test CMB data for isotropy.  

The wave function $\psi_{m}^{k}(l)$ is a linear transformation of the amplitudes $a_{lm}$ which expresses them as products of vectors and representations $l$.  The power tensor $A_{ij}$ is quadratic in the wave functions and $a_{lm}$.  {\it Invariants} of the power tensor are its eigenvalues, the singular values of $\psi$.  The statistical distribution of these invariants is a test of randomness and isotropy of the the power tensor $l$ by $l$.  The {\it power entropy} serves as a robust statistic. The power entropy is independent of the usual power, and also independent of the eigenvalues of $A_{ij}$, which contain information on the orientation of multipoles.  To assess multipole orientations, we collected the principal axis of $A_{ij}$, which is the eigenvector with largest eigenvalue.

Both the power entropy and the axial vector distributions show features that are not consistent with isotropy.  Fig. \ref{fig:SinValPVal2.eps} summarizes some results of the ILC-full sky study. The figure repeats Fig. \ref{fig:ILCSingValsVSl.eps} highlighting the exceptional cases. Cases of very low power entropy are joined by a short thing line.  Cases of very high alignment of the principal axis with the quadrupole axis are shown by a longer thick gray line. There are 12 independent cases having 5\%  probability or lower in an isotropic distribution, with 2 coincidences showing {\it both} exceptional power entropy and alignment.  Ignoring the double coincidences, there are 98 independent trials in examining $2<l<49$ twice.  The total probability to see the data's values in uncorrelated isotropic CMB maps is of order $10^{-3}$. 

Our study substantially extends previous studies that find low $l$ multipoles of current CMB maps are highly unlikely on statistical grounds.  Unlike methods that produce a large number of vector factors for each $l$, our wave function method identifies a single invariant axis for each $l$, and 3 independent invariants for each $l$. The axial quantities are remarkably aligned with the direction of Virgo in rather systematic fashion for many $l$.  CMB data show 7 total axes aligned over the range $1<l<50$, with each case having independent probability at order $5\%$ or smaller.  The values of $l$= 6, 16, 17, 30, 34, 40  have anomalously non-random power eigenvalue distributions giving exceptionally lower power entropy.  Multipoles with $l=$ 3, 9, 16, 21, 40, 43 have anomalous alignment with the quadrupole. It is no longer possible to restrict anomalous CMB statistics to the relative orientation or power seen in the dipole, quadrupole and octupole cases. Our study shows, for the first time, that the alignment of the 
multipoles with an axis pointing in the direction of Virgo 
is much more pervasive 
and not just confined to low $l$ values. 

To explore the origin of these features, including possible foreground effects, we repeated many calculations using sky-masked data sets.  The data shows evidence for systematic differences between the Northern and Southern regions of the sky.  Masking out the galactic plane does not eliminate signals of anisotropy seen in the full sky studies.  As consistency checks, the anisotropies of CMB plus simulated synchrotron emission contamination from the plane of the galaxy are detected by our methods in just the spatial regions where they are simulated. The observed anisotropies cannot readily be explained away by appealing to galactic foreground contamination. 
 
\medskip 
{\bf{ Acknowledgement}}
We acknowledge the use of Legacy Archive for Microwave Background Data Analysis. Pramoda acknowledges CSIR, India for finacial support under the research grant CSIR-SRF-9/92(340)/2004-EMR-I.  
Ralston is supported in part under DOE Grant Number
DE-FG02-04ER14308.

\section{Appendix} 

Here we relate the factorization wave function $\psi_{m}^{k}$ to covariant canonical forms of the multipoles $a_{lm}$. 

In the theory of angular momentum one conventionally decomposes {\it products} of representations of dimensions $j_{1}$, $j_{2}$ into sums of states $\ket{JM}$, where $|j_{1}-j_{2}|<J<j_{1}+j_{2}$.  It is also possible to {\it divide} the large representation $\ket{JM}$ into products of smaller representations.  Let $\Psi_{M}^{J} =\bracket{JM}{\Psi}$ be the matrix elements of a state
$\ket{\Psi}$ transforming on representation $J$. We write \ba \Psi_{M}^{J} \ra \psi_{  m_{1} m_{2}}^{j_{1} j_{2} }= \Gamma_{JM}^{j_{1} m_{1}j_{2}m_{2}*} \Psi_{M}^{J}, \ea where the Clebsh-Gordan coefficients are \ba \Gamma_{JM}^{j_{1} m_{1}j_{2}m_{2}*} = \bracket{j_{1}m_{1}j_{2}m_{2}}{JM}.  \nn \ea 

The Clebsch's dividing angular momentum $J$ in products of $j_{1}=1$ and $j_{2}=J$ are simply the angular momentum generators $\vec J$ cast into the angular momentum basis. The relation of $\psi_{m}^{k}$ to 
$\psi_{  m_{1} m_{2}}^{j_{1} j_{2} }$ uses $\ket{\Psi} \ra \ket{a(l)}$, sets $m=m_{1}$, converts one Cartesian index $k$ into angular momentum $m_{2}$, uses $j_{1}=1$ and $j_{2}=J=i$.   

There exist a large number of decompositions for different choices of factorization.  Each particular choice has a privileged status when viewed as the unique way to factor the multiplet into products of smaller spaces.  There is no principle upon which to choose the smaller spaces, explaining the remark that no preferred canonical form exists for high-rank tensors.  The relation of $\psi_{m}^{k}$ to the canonical form of Copi et al can be viewed as coming from two steps.  First divide $\ket{a} \ra (j_{1} =1 )\otimes (j_{1} =1 )\otimes..(j_{2} =1 )...\otimes.(j_{l} =1 ) $, to develop the Maxwell multipoles.  Next find a unique preferred vector which multiplies a  tensor of rank $l$ made from traceless symmetric products of the other vectors. That vector corresponds to $\psi_{m}^{k}$.  
Alternatively $\psi_{m}^{k}$ can simply be written down (Ralston and Jain 2004) as the vector available from symmetry. 
 
\bigskip
\noindent
 \bf{\large  References}

\begin{itemize}

\item[] Abramo, L. R.,  Sodre Jr., L., Wuensche, C. A., 2006,
Phys. Rev. D74, 083515

\item[]  Armendariz-Picon, C., 2004, 
JCAP 0407, 007

\item[]  Armendariz-Picon, C., 2006, 
JCAP 0603, 002

\item[] Battye, R. A.,  Moss, A., 2006, Phys. Rev. D74, 04130

\item[] Bernui, A., Villela, T., Wuensche, C. A., Leonardi, R.,
 Ferreira, I., 2006, Astron. Astrophys. 454, 409

\item[] Bernui, A., Mota, B., Reboucas, M. J., Tavakol, R., 
2007, Astron. Astrophys. 464, 479

\item[]  Bietenholz, M. F.,   Kronberg, P. P., 1984,   ApJ,
 {287}, L1-L2 

\item[] Bielewicz,  P., Gorski,  K. M., Banday, A.J., 2004, 
MNRAS, 355, 1283

\item[] Bielewicz, P., Eriksen, H. K., Banday, A. J., Gorski, K. M.,
 Lilje,  P. B., 2005, ApJ 635, 750

\item[]
 Birch, P., 1982,   Nature,  {298}, 451

\item[] Buniy, R. V., Berera, A., Kephart, T. W., Phys. Rev. D73, 063529 

\item[]  Campanelli, L., Cea, P., Tedesco, L., 2007, arXiv:0706.3802

\item[] Cline, J. M.,  Crotty, P.,  Lesgourgues, J., 2003, JCAP 0309, 010,
astro-ph/0304558;  

\item[] Contaldi, C. R.,  Peloso, M,  Kofman, L.,  Linde, A., 2003, 
JCAP 0307, 002, astro-ph/0303636  

\item[] Copi, C. J., Huterer, D., Schwarz, D. J,  Starkman, G. D., 2006,
MNRAS 367, 79

\item[] Copi, C. J., Huterer, D., Schwarz, D. J,  Starkman, G. D., 2007,
  Phys.\ Rev.\  D75, 023507 
  
\item[] da Silva. V., Lim, L-H., 2006, math-0607647
 
\item[] de Oliveira-Costa, A.,  Tegmark, M., Zaldarriaga, M., 
Hamilton, A., 2004, Phys. Rev. D 69, 063516,
astro-ph/0307282

\item[]  de Oliveira-Costa, A.,  Tegmark, M., 2006, 
Phys. Rev. D74, 023005

\item[]  Dennis, M. R., 2005, J. Phys. A 38, 1653,
arXiv:math-ph/0410004

\item[] Donoghue E. P.,  Donoghue, J. F., 2005, Phys. Rev. D71, 043002

\item[] Efstathiou, G., 2003, MNRAS  346, L26,
astro-ph/0306431

\item[] Eriksen, H. K. {\it et al}, 2004, ApJ  605, 14

\item[] Eriksen, H. K. {\it et al}, 2007,  ApJ  656,  641

\item[] Freeman, P. E., Genovese, C.R., Miller, C.J., Nichol, R.C., 
Wasserman, L., 2006, ApJ, 638, 1

\item[] Gaztanaga, E., Wagg, J., Multamaki, T., Montana, A.,  Hughes, D. H., 2003, 
MNRAS 346, 47, astro-ph/0304178 

\item[] Giardino et al. (2002) {\it A \& A },  387, 82

\item[]  Gordon, C., Hu,  W., Huterer, D.,  Crawford, T. 2005, 
Phys. Rev. D72, 103002

\item[] Gumrukcuoglu, A. E., Contaldi, C. R., Peloso, M., 2006, 
astro-ph/0608405

\item[] Gumrukcuoglu, A. E., Contaldi, C. R., Peloso, M., 2007, 
arXiv:0707.4179

\item[] Hajian, A., Souradeep,  T., Cornish,  N. J., 2004, 
ApJ 618, L63

\item[]  Hajian, A.,  Souradeep, T., 2006, Phys. Rev. D74, 123521

\item[] Helling, R. C., Schupp, P., Tesileanu T., 2007, Phys. Rev. D74,
063004

\item[] Hunt, P., Sarkar, S., 2004, Phys. Rev. D70, 103518

\item[] 
Hinshaw, G. {\it  et al.}, 2003, Astrophys. J., Suppl. Ser.  148, 63

\item[] Hinshaw, G. {\it et al}, 2007, ApJ Suppl 170, 288 

\item[]  Hivon, E., {\it et al}, 2002,  ApJ,  567, 2
\item[] 
Hutsem\'{e}kers, D., 1998 {\it A \& A},   332, 41 

\item[]  Hutsem\'{e}kers, D., Lamy, H., 2001,  {\it A \& A},  {367}, 381 

\item[] Inoue, K. T., Silk, J., 2006, ApJ 648, 23 

\item[] Jain, P.,  Narain, G.,   Sarala, S., 2004, {\it MNRAS},  347, 394 

\item[] Jain, P,  Ralston, J. P., 1999,  Mod. Phys. Lett. A14, 417 

\item[]  Jain, P.,  Sarala, S., 2006, J. Astrophysics and Astronomy,  27, 443 

\item[] Land, K.,  Magueijo, J., 2005, Phys. Rev. D72, 101302 

\item[] Land, K.,  Magueijo, J., 2006, MNRAS 367, 1714 

\item[] Land, K., Magueijo, J., 2007, MNRAS 378, 153 

\item[] Katz, G., Weeks,  J., 2004, Phys. Rev. D70, 063527

\item[] Kendall, D. G., Young, A. G., 1984,  MNRAS,  {207}, 637 

\item[] Kesden, M. H., Kamionkowski, M., Cooray, A., 2003,  
Phys. Rev. Lett. 91,  221302, astro-ph/0306597  

\item[]  Kogut, A. {\it et al}, 2003, Astrophys. J. Suppl.,  148, 161

\item[] Koivisto, T., Mota, D. F., 2006, Phys. Rev. D73, 083502

\item[] Magueijo, J.,  Sorkin, R. D., 2007, MNRAS Lett. 377, L39

\item[] Moffat, J. W., 2005, JCAP 0510, 012

\item[] 
Naselsky, P. D., Verkhodanov, O. V.,  Nielsen, M. T. B., 2007,
arXiv:0707.1484

\item[]   Nodland, B., Ralston, J.~P.
  Phys.\ Rev.\ Lett.\  {\bf 78}, 3043 (1997)

\item[] Pereira, S., Pitrou, C.,  Uzan, J.-P., 2007, arXiv:0707.0736

\item[] Phinney, E. S.,  Webster, R. I., 1983, {\it Nature}
{ 301}, 735 

\item[] Prunet, S.,  Uzan, J.-P.,  Bernardeau, F., Brunier, T., 2005,
Phys. Rev. D71, 083508

\item[] Rakic, A., Rasanen, S., Schwarz,  D. J., 2006, MNRAS 
369, L27

\item[]
     Ralston, J. P.,  Jain, P., 1999,
     Int. J. Mod. Phys. {\bf{ D 8}}, 537. 

\item[]  Ralston, J. P.,  Jain, P., 2004,  Int. J. Mod. Phys., 
{\bf{ D13}}, 1857 

\item[] Rodrigues, D. C., 2007, arXiv:0708.1168

\item[] Saha, R., Jain, P. and Souradeep, T., 2006, ApJ Lett.,
 645, L89

\item[]  Schwarz, D. J., Starkman, G. D., Huterer, D.,  Copi, C. J., 2004,
Phys. Rev. Lett.  93, 221301 

\item[] Slosar, A.,  Seljak, U., 2004, Phys. Rev. D70, 083002

\item[] Tegmark, M.,  de  Oliveira-Costa, A. and Hamilton, A., 2003,
Phys. Rev. D  68, 123523 

\item[] Vale, C., 2005, astro-ph/0509039

\item[] Von Neumann, J., 1932, {\it The Mathematical Foundations of Quantum Mechanics}  translated by R. T. Beyer,   (Princeton University Press 1996)

\item[] Weeks, J. R., 2004, astro-ph/0412231

\item[] Wiaux, Y., Vielva, P.,  Martinez-Gonzalez, E.,  Vandergheynst, P.,
2006, Phys. Rev. Lett. 96, 151303 
\end{itemize}

\end{document}